\documentclass[3p]{elsarticle}

\usepackage{etex}

\usepackage{amssymb}
\usepackage{graphicx}
\usepackage{caption}
\usepackage{subcaption}
\usepackage{hyperref}
\usepackage{amsmath}
\usepackage{z-eves}
\usepackage[linesnumbered,ruled,vlined]{algorithm2e}
\usepackage{verbatim}
\usepackage{longtable}
\usepackage{xcolor}
\usepackage{soul}

\usepackage{pgf}
\usepackage{tikz}
\usetikzlibrary{arrows,automata,shapes}
\usepackage{pgfplots}
\usepackage{coqdoc}

\usepackage{pgf}
\usepackage{tikz}
\usetikzlibrary{arrows,automata,shapes}
\usepackage{pgfplots}

\usepackage[linesnumbered,ruled,vlined]{algorithm2e}
\usepackage{z-eves}
\usepackage{cspsymb}
\usepackage{acronym}

\usepackage{listings}
\begin{document}


\begin{frontmatter}

\title{Modelling and testing timed data-flow reactive systems in Coq\\ from controlled natural-language requirements}
\author[cin]{Gustavo Carvalho}
\ead{ghpc@cin.ufpe.br}

\author[cin]{Igor Meira}
\ead{iam2@cin.ufpe.br}

\address[cin]{Universidade Federal de Pernambuco -- Centro de Inform\'atica, 50740-560, Brazil}

\begin{abstract} 
Data-flow reactive systems (DFRSs) are a class of embedded systems whose inputs and outputs are always available as signals. Input signals can be seen as data provided by sensors, whereas the output data are provided to system actuators. In previous works, verifying properties of DFRS models was accomplished in a programmatic way, with no formal guarantees, and test cases were generated by translating theses models into other notations. Here, we use Coq as a single framework to specify and verify DFRS models. Moreover, the specification of DFRSs in Coq is automatically derived from controlled natural-language requirements. Property verification is defined in both logical and functional terms. The latter allows for easier proof construction. Tests are generated with the support of the QuickChick tool. Considering examples from the literature, but also from the aerospace industry (Embraer), our testing strategy was evaluated in terms of performance and the ability to detect defects generated by mutation; within 8 seconds, we achieved an average mutation score of 75.80\%.
\end{abstract}


\end{frontmatter}



\section{Introduction}\label{sec:intro}

Over the years, we have been building a society that is highly dependent on software. In many situations, the software is part of a safety-critical system, and all sorts of failures shall be minimised, since they might be life-threating or impose significant financial losses. Therefore, high trustworthiness levels are typically required for such systems.

Modelling and formal verification are strategies employed to increase system reliability. Creating models promotes a better comprehension and a more precise description of the expected behaviour. Formal verification brings certainty about properties being preserved. For instance, considering the avionics industry, in~\cite{cofer:nfm_report}, three cases studies are reported illustrating the use of different classes of formal methods (theorem proving, model checking, and abstract interpretation) to meet reliability levels defined by the standard DO-178C (Software Considerations in Airbone Systems and Equipment Certification). When formal verification is not possible or feasible, testing becomes essential since it can unveil scenarios where implementations do not work as expected.

\subsection{Model-based testing}

In a model-based testing (MBT) strategy, test cases are derived from models, making the testing process more agile, less susceptible to errors, and less dependent on human interaction. This goal is usually reached by means of automatic generation (and execution) of test cases, besides automatic generation of test data, from specification models. 

Here, we focus on models of data-flow reactive systems (DFRS): a class of embedded systems whose inputs and outputs are always available as signals. Additionally, the system behaviour might be time dependent. Models of DFRSs are fully explained in~\cite{carvalho:dfrs_journal}. They have been used to model examples both from the literature and the industry. In this previous work, we also show that these models can be seen as timed input-output transition systems, but, being more abstract, enable automatic extraction from system-level specifications in a controlled natural language, which is an important aspect as discussed in what follows.

Despite the benefits of MBT, those who are not familiar with the models syntax and semantics may be reluctant to adopt theses formalisms. This is particularly more evident when considering formal MBT strategies, when formal models of the system expected behaviour need to be developed. Moreover, most of these models are not available in the very beginning of the project, when usually natural-language requirements are available. One possible alternative to overcome these limitations is to employ NLP techniques to derive the required models from natural-language specifications automatically.

\subsection{Natural-language processing}

The demand of stating the desired system behaviour using formal models may sometimes be an obstacle for adopting formal MBT techniques, despite all its benefits. The model notations may be not easy to interpret by, for instance, aerospace and automotive development engineers. Hence, a specialist (usually mathematicians, logicians, computer engineers and scientists) is required when such languages, and their corresponding techniques, are used in business contexts. Furthermore, most of these models are not yet available in the very beginning of the system development project.

As previously said, one possible alternative to overcome these limitations is to provide means for deriving specification models automatically from the already existing documentation, in particular, natural-language requirements. In this sense, NLP techniques can be helpful. If formal models are derived from natural-language requirements, besides applying MBT techniques, one can reason formally about properties of specifications that can be difficult to analyse by means of manual inspection, such as inconsistency and incompleteness.

Typically, there is a trade-off concerning the application of NLP in MBT. Some studies are able to analyse a broad range of sentences, whereas others rely on controlled versions of natural language (CNL). The works that adopt the former approach usually depend on a higher level of user intervention to derive models and to generate test cases. Differently, the restrictions imposed by a CNL might allow a more automatic approach when generating models and test cases. Ideally, a compromise between these two possibilities should be sought to provide a useful degree of automation along with a natural-language specification feasible to be used in practice.

In this work, seeking for automation, we adopt a CNL for describing the system requirements. In~\cite{carvalho:dfrs_journal}, we provide a comprehensive explanation of how models of DFRSs can be automatically derived from SysReq-CNL, a CNL specially designed for editing requirements of data-flow reactive systems.

\subsection{The NAT2TEST$_{Coq}$ strategy}

Considering a characterisation of DFRSs in the Coq proof assistant~\cite{bertot:coqart}, we expand here the limits of the NAT2TEST strategy~\cite{carvalho:tool}, which is devised to generate test cases from natural-language requirements. The specialisations of this strategy allow for exploring the benefits of different formal techniques, such as SMT solving, model and refinement checking, and simulation.

The specialisation proposed in this paper (hereafter, named NAT2TEST$_{Coq}$) is the first one based on an interactive proof assistant (Coq), which also brings new benefits and possibilities. For instance, in our previous efforts, verification of consistency properties of DFRS models was accomplished in a programmatic way, decoupled from the formal consistency definition. Here, we use the Coq proof assistant as a single framework to both specify and verify properties of DFRS models. Property verification is defined in both logical and functional terms. The latter allows for easier proof construction, besides contributing to the development of correct-by-construction tools (via extraction of Haskell/Ocaml code from the functional definitions), which is currently a future work.

Furthermore, tests are generated via the QuickChick tool~\cite{pierce:quickchick}. Considering examples from the literature, but also from the aerospace industry (Embraer\footnote{Embraer website: \url{https://embraer.com/global/en}}), our testing strategy was evaluated in terms of performance and the ability to detect defects generated by mutation; within 8 seconds, we achieved an average mutation score of 75.80\%.

Therefore, the main contribution of this work is a single and consistent framework, based on Coq, for modelling, verifying and testing timed reactive systems from natural-language requirements; besides providing empirical evidence on the quality of our test generation strategy via mutant-based strength analysis. The remainder of this paper is organised as follows. Section~\ref{sec:bg} discusses the main concepts related to this work (the NAT2TEST strategy, data-flow reactive systems, interactive theorem proving, and property-based testing). Sections~\ref{sec:sdfrs_coq} and~\ref{sec:edfrs_coq} present our logical and functional characterisation of symbolic and expanded data-flow reactive systems, respectively. Section~\ref{sec:quickchick} addresses test generation, based on the QuickChick tool. Our empirical analyses are presented in Section~\ref{sec:empirical}. Finally, Section~\ref{sec:conclusion} concludes this paper by discussing related and future work.

\section{Background}\label{sec:bg}

Now, we present the foundational concepts related to this work: the NAT2TEST strategy (Section~\ref{sec:nat2test}), DFRS models (Section~\ref{sec:dfrs}), Coq (Section~\ref{sec:coq}), and property-based testing (Section~\ref{sec:pbt}).

\subsection{The NAT2TEST strategy}\label{sec:nat2test}

\begin{figure}[t]
	\centering
	\includegraphics[width=0.75\textwidth]{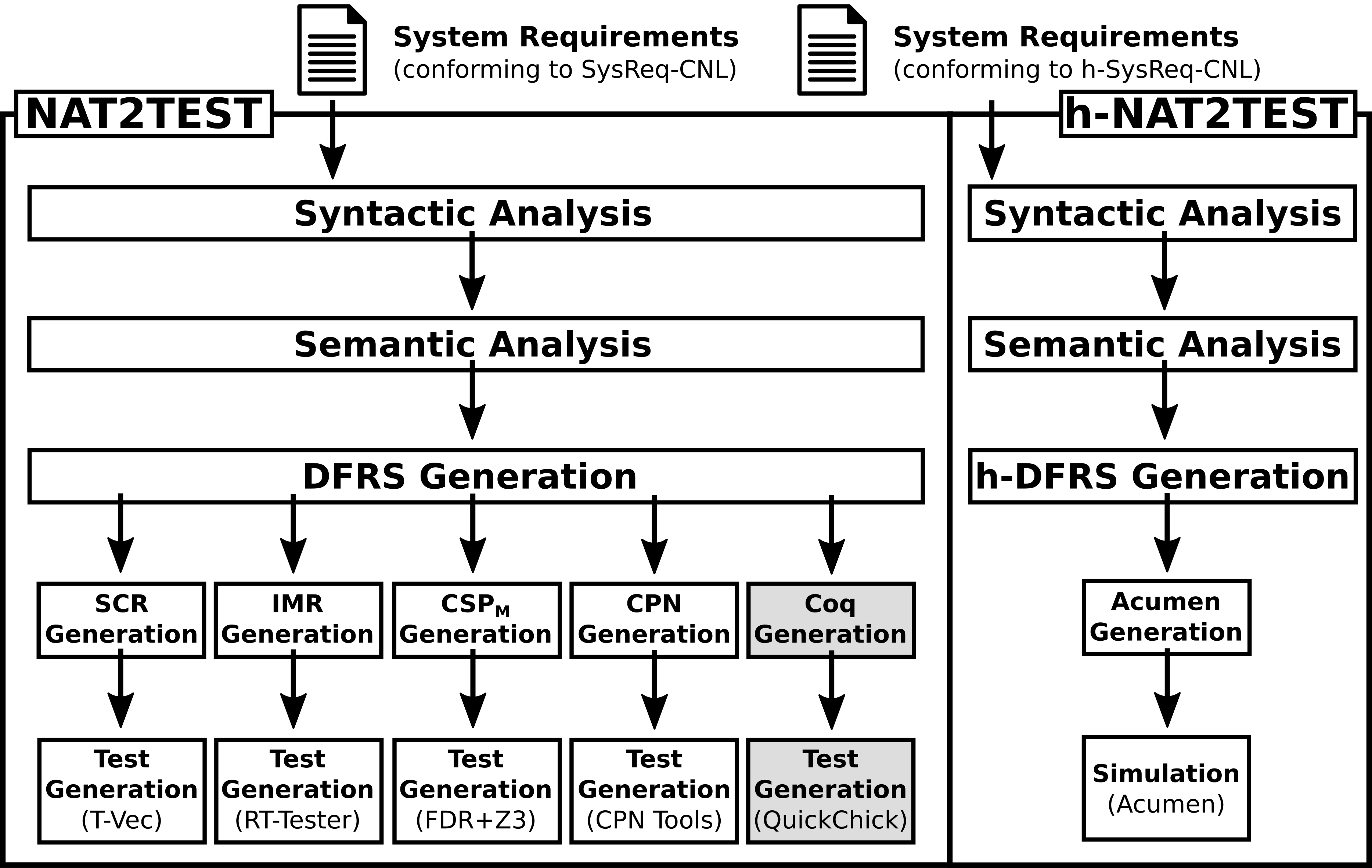}
	\caption{NAT2TEST -- a strategy for generating test cases based on different formalisms}
	\label{fig:nat2test_family}
\end{figure}

The NAT2TEST strategy is tailored to generate tests for timed data-flow reactive systems, considering different internal and hidden formalisms (see Figure~\ref{fig:nat2test_family}). This test-generation strategy comprises a number of phases. The three initial phases are fixed: (1) syntactic analysis, (2) semantic analysis, and (3) DFRS generation; the remaining phases depend on the internal formalism chosen.

\paragraph{Syntactic analysis} In this work, requirements are written according to a CNL based on English: the SysReq-CNL, specially designed for editing requirements of data-flow reactive systems. The first phase of the NAT2TEST strategy is responsible for verifying whether the requirements are in accordance with the SysReq-CNL grammar. For each valid requirement, its corresponding syntax tree is identified.

As a running example, we consider a Vending Machine~(VM) (adapted from the coffee machine presented in~\cite{larsen:rtioco}). Initially, the VM is in an \emph{idle} state. When it receives a coin, it goes to the \emph{choice} state. After inserting a coin, when the coffee option is selected, the system goes to the \emph{weak} or \emph{strong} coffee state. If coffee is selected within 30 seconds after inserting the coin, the system goes to the \emph{weak coffee} state. Otherwise, it goes to the \emph{strong coffee} state. The time required to produce a weak coffee is also different from that of a strong coffee; the former is produced within 10 to 30 seconds, whereas the latter within 30 to 50 seconds. After producing coffee, the system returns to the idle state.

The following sentence exemplifies a requirement (REQ001) that adheres to the SysReq-CNL grammar: \emph{When the system mode is idle, and the coin sensor changes to true, the coffee machine system shall: reset the request timer, assign choice to the system mode}.

\paragraph{Semantic analysis} In the second phase, the requirements are semantically analysed using the case grammar theory~\cite{fillmore:CG}. In this theory, a sentence is not analysed in terms of the syntactic categories or grammatical functions, but in terms of the semantic (thematic) roles played by each word/group of words in the sentence. Therefore, for each syntax tree, the group of words that correspond to a thematic role is identified. The collection of thematic roles for a requirement is called a requirement frame.

Table~\ref{tab:trExampleIntro} shows the requirement frame for REQ001. We note that the thematic roles are grouped into conditions and actions. The roles that appear in actions are the following: \emph{Action} -- the action performed if the conditions are satisfied; \emph{Agent} -- entity who performs the action; \emph{Patient} -- entity who is affected by the action; and \emph{To Value} -- the patient value after action completion. Similar roles appear in conditions.

\begin{table}[htb]
	\centering
	\caption{\small{Example of requirement frame for REQ001 (VM)}}
	\label{tab:trExampleIntro}
	\normalsize{
		\begin{tabular}{llll} \hline\hline
			\multicolumn{4}{l}{\textbf{Condition \#1} - Main Verb (Condition Action): \emph{is}} \\ \hline
			Condition Patient: & \emph{the system mode} & Condition From Value: & -- \\ \hline
			Condition Modifier: & -- & Condition To Value: & \emph{idle} \\ \hline\hline
			\multicolumn{4}{l}{\textbf{Condition \#2} - Main Verb (Condition Action): \emph{changes}} \\ \hline
			Condition Patient: & \emph{the coin sensor} & Condition From Value: & -- \\ \hline
			Condition Modifier: & -- & Condition To Value: & \emph{true} \\ \hline\hline
			\multicolumn{4}{l}{\textbf{Action} - Main Verb (Action): \emph{reset}} \\ \hline
			Agent: & \emph{the coffee machine system} & To Value: & -- \\ \hline
			Patient: & \emph{the request timer} & & \\ \hline\hline
			\multicolumn{4}{l}{\textbf{Action} -- Main Verb (Action): \emph{assign}} \\ \hline
			Agent: & \emph{the coffee machine system} & To Value: & \emph{choice} \\ \hline
			Patient: & \emph{the system mode} & & \\ \hline\hline
		\end{tabular}    
	}
\end{table}

\paragraph{DFRS generation} Afterwards, the third phase derives DFRS models -- an intermediate formal characterisation of the system behaviour from which other formal notations can be derived. The possibility of exploring different formal notations allows analyses from several perspectives, using different languages, tools, and techniques. Besides that, it makes our strategy extensible. Models of DFRSs are explained in the following section. For a comprehensive explanation of how DFRS models are derived from requirement frames, we refer to~\cite{carvalho:dfrs_journal}.

\paragraph{Test generation} Test generation is achieved by translating DFRS models into internal and hidden formalisms. In what follows, we list the possibilities currently supported by the NAT2TEST strategy.

\begin{itemize}
	\item NAT2TEST$_{SCR}$: based on \emph{Software Cost Reduction} -- SCR~\cite{heninger:SCRFirst} (more details in \cite{carvalho:scp})
	\item NAT2TEST$_{IMR}$: based on \emph{Internal Model Representation} -- IMR~\cite{peleska:rttester} (more details in \cite{carvalho:rttester})
	\item NAT2TEST$_{CSP}$: based on \emph{Communicating Sequential Processes} -- CSP~\cite{roscoe:new_book} (more details in \cite{carvalho:csptio})
	\item NAT2TEST$_{CPN}$: based on \emph{Coloured Petri Nets} -- CPN~\cite{kurt:cpn} (more details in \cite{carvalho:cpn_journal})
	\item NAT2TEST$_{Coq}$: based on \emph{Coq}~\cite{bertot:coqart} -- the main contribution of this paper
\end{itemize}

As said before, exploring different formal notations allows test generation from several perspectives, using different languages, tools, and techniques. The strategies NAT2TEST$_{SCR}$ and NAT2TEST$_{IMR}$ generate test cases with the support of commercial testing tools: T-VEC\footnote{T-VEC website: \url{https://www.t-vec.com/}} and RT-Tester\footnote{RT-Tester website: \url{https://www.verified.de/products/rt-tester/}}, respectively. Differently, the NAT2TEST$_{CSP}$ strategy reuses a general purpose refinement checker (FDR\footnote{FDR website: \url{https://www.cs.ox.ac.uk/projects/fdr/}}) and SMT solver (Z3\footnote{Z3 website: \url{https://github.com/Z3Prover/z3}}) to deliver a formal and sound testing theory. Scalability is a known issue of this specialisation of the NAT2TEST strategy. Differently, the NAT2TEST$_{CPN}$ strategy aims at efficiency by generating test cases via random simulation of CPN models. Table~\ref{tab:nat2test_family} summarises the languages, tools and techniques considered by the aforementioned specialisations in order to generate test cases.

\begin{table}[htb]
	\centering
	\caption{\small{Specialisations of the NAT2TEST strategy -- techniques for generating test cases}}
	\label{tab:nat2test_family}
	\normalsize{
		\begin{tabular}{l||c|c|c} \hline\hline
			& \textbf{Internal formalism} & \textbf{Applied techniques} & \textbf{Tools integration} \\ \hline\hline
			\textbf{NAT2TEST$_{SCR}$} & SCR & SMT solving & T-VEC \\ \hline\hline
			\textbf{NAT2TEST$_{IMR}$} & IMR & SMT solving & RT-Tester \\ \hline\hline
			\textbf{NAT2TEST$_{CSP}$} & CSP$_{\text{M}}$ & Model checking + SMT solving & FDR + Z3\\ \hline\hline
			\textbf{NAT2TEST$_{CPN}$} & CPN & Simulation & CPN Tools\\ \hline\hline
			\textbf{NAT2TEST$_{Coq}$} & Coq & Property-based testing & Coq + QuickChick\\ \hline\hline
		\end{tabular}    
	}
\end{table}

The NAT2TEST$_{Coq}$ strategy distinguishes itself by integrating the NAT2TEST strategy with a technique not explored so far (proof assistants), and generating test cases via property-based testing. Now, besides test generation, it is also possible to develop proof scripts in Coq for proving relevant properties.

\paragraph{Other extensions} In~\cite{carvalho:environment}, we extend our CNL to allow the specification of environment restrictions and, thus, how the system interacts with its surrounding environment. This extension has only been incorporated into the NAT2TEST$_{CSP}$ specialisation. In this way, unrealistic interactions between the system and the environment are not considered when generating test cases via FDR + Z3.

In~\cite{carvalho:nusmv}, we allow the specification in natural language of system properties (in the style of temporal logic). These properties, along with the system requirements, are translated into CTL formulae and NuSMV models, respectively. With the aid of the NuSMV model checker~\cite{cimatti:nusmv}, it is possible to assess whether the specified properties are satisfied by the NuSMV models. Here, test generation is not the ultimate goal, but model checking requirements.

Finally, in~\cite{carvalho:hybrid}, we discuss a vertical adaptation of the NAT2TEST strategy in order to simulate hybrid systems (featuring the integration of discrete and continuous behavioural aspects) also from requirements adhering to a CNL. Therefore, in this work we revisit each phase of the NAT2TEST strategy, now considering the h-SysReq-CNL (an extension of the SysReq-CNL where it is possible to define differential equations) and a hybrid version of DFRS models (h-DFRS). Simulation is enabled by translating h-DFRS models into Acumen~\cite{walid:acumen}, which is a language and tool for the specification and simulation of hybrid systems.

\subsection{Data-flow reactive systems}\label{sec:dfrs}

A data-flow reactive system (DFRS) is an embedded system whose inputs and outputs are always available, as signals. The input signals can be seen as data provided by sensors, whereas the outputs are data provided to actuators. A DFRS can also have internal timers, which are used to trigger time-based behaviour. There are two models of DFRSs: a symbolic (s-DFRS) and an expanded (e-DFRS) one. Briefly speaking, the former comprises an initial state, along with functions that describe the system behaviour (how the system state might evolve). Differently, an e-DFRS represents the system behaviour as a state-based machine; it can be seen as an expansion of its symbolic counterpart by applying the s-DFRS functions to its initial state, but also to the new reachable states.

As a running example, we consider the VM (presented previously). In this example, we have two input signals related to the coin sensor (\emph{sensor}) and the coffee request button (\emph{request}). A \emph{true} value means that a coin was inserted or that the coffee request button was pressed, respectively. There are two output signals: one related to the system mode (\emph{mode}) and another to the vending machine output (\emph{output}). The values communicated by these signals reflect the system's possible modes (\emph{choice} $\mapsto 0$, \emph{idle} $\mapsto 1$, \emph{strong coffee} $\mapsto 2$, and \emph{weak coffee} $\mapsto 3$) and the possible outputs (\emph{strong} $\mapsto 0$, and \emph{weak} $\mapsto 1$). The VM has just one timer: the \emph{request} timer, which is used to register the moments when a coin is inserted, when the coffee request button is pressed, and when the coffee is produced.

\begin{figure}[!h]
	\centering
	\begin{subfigure}{.5\textwidth}
		\centering
		\begin{tikzpicture}
		\tikzstyle{hidden} = [coordinate]
		\tikzstyle{every label}=[label position=left]
		\node [hidden] at (0,0) [label={\scriptsize{false}}] {};
		\node [hidden] at (0,1) [label={\scriptsize{true}}] {};
		\tikzstyle{every label}=[label position=below]
		\node [hidden] at (0.2,0) [label={\scriptsize{2}}] {};
		\node [hidden] at (0.5,0) [label={\scriptsize{5}}] {};
		\node [hidden] at (3.0,0) [label={\scriptsize{30}}] {};
		
		\draw[dotted,color=gray] (0,0) grid (3.0,1.0);
		\draw[thin,->] (0,0) -- (3.2,0) node[right] {$time (s)$};
		\draw[thin,-] (0,0) -- (0,1) node[rotate=90,xshift=-3.5ex,yshift=7.0ex] {$sensor$};	
		\draw[very thick] plot coordinates {(0,0) (0.2,0) (0.2,1) (0.5,1) (0.5,0) (3,0)};
		\end{tikzpicture}
		\begin{tikzpicture}
		\tikzstyle{hidden} = [coordinate]
		\tikzstyle{every label}=[label position=left]
		\node [hidden] at (0,0) [label={\scriptsize{false}}] {};
		\node [hidden] at (0,1) [label={\scriptsize{true}}] {};
		\tikzstyle{every label}=[label position=below]
		\node [hidden] at (1.2,0) [label={\scriptsize{12}}] {};
		\node [hidden] at (1.6,0) [label={\scriptsize{16}}] {};
		\node [hidden] at (3.0,0) [label={\scriptsize{30}}] {};
		
		\draw[dotted,color=gray] (0,0) grid (3.0,1.0);
		\draw[thin,->] (0,0) -- (3.2,0) node[right] {$time (s)$};	
		\draw[thin,-] (0,0) -- (0,1) node[rotate=90,xshift=-3.5ex,yshift=7.0ex] {$request$};	
		\draw[very thick] plot coordinates {(0,0) (1.2,0) (1.2,1) (1.6,1) (1.6,0) (3,0)};
		\end{tikzpicture}
		\begin{tikzpicture}
		\tikzstyle{hidden} = [coordinate]
		\tikzstyle{every label}=[label position=left]
		\node [hidden] at (0,0) [label={\scriptsize{0}}] {};
		\node [hidden] at (0,1) [label={\scriptsize{1}}] {};
		\tikzstyle{every label}=[label position=below]
		\node [hidden] at (2.6,0) [label={\scriptsize{26}}] {};
		\node [hidden] at (3.0,0) [label={\scriptsize{30}}] {};
		
		\draw[dotted,color=gray] (0,0) grid (3.0,1.0);
		\draw[thin,->] (0,0) -- (3.2,0) node[right] {$time (s)$};	
		\draw[thin,-] (0,0) -- (0,1) node[rotate=90,xshift=-3.5ex,yshift=7.0ex] {$output$};	
		\draw[very thick] plot coordinates {(0,0) (2.6,0) (2.6,1) (3.0,1)};
		\end{tikzpicture}				
	\end{subfigure}%
	\begin{subfigure}{.5\textwidth}
		\centering
		\begin{tikzpicture}
		\tikzstyle{hidden} = [coordinate]
		\tikzstyle{every label}=[label position=left]
		\node [hidden] at (0,0) [label={\scriptsize{0}}] {};
		\node [hidden] at (0,1) [label={\scriptsize{1}}] {};
		\node [hidden] at (0,2) [label={\scriptsize{2}}] {};
		\node [hidden] at (0,3) [label={\scriptsize{3}}] {};
		\tikzstyle{every label}=[label position=below]
		\node [hidden] at (0.2,0) [label={\scriptsize{2}}] {};
		\node [hidden] at (1.2,0) [label={\scriptsize{12}}] {};
		\node [hidden] at (2.6,0) [label={\scriptsize{26}}] {};
		\node [hidden] at (3.0,0) [label={\scriptsize{30}}] {};			
		
		\draw[dotted,color=gray] (0,0) grid (3.0,3.0);
		\draw[thin,->] (0,0) -- (3.2,0) node[right] {$time (s)$};	
		\draw[thin,-] (0,0) -- (0,3) node[rotate=90,xshift=-11.0ex,yshift=6.0ex] {$mode$};	
		\draw[very thick] plot coordinates {(0,1) (0.2,1) (0.2,0) (1.2,0) (1.2,3) (2.6,3) (2.6,1) (3.0,1)};
		\end{tikzpicture}
	\end{subfigure}
	\caption{Example of signals for the vending machine}
	\label{fig:vm-signals}	
\end{figure}
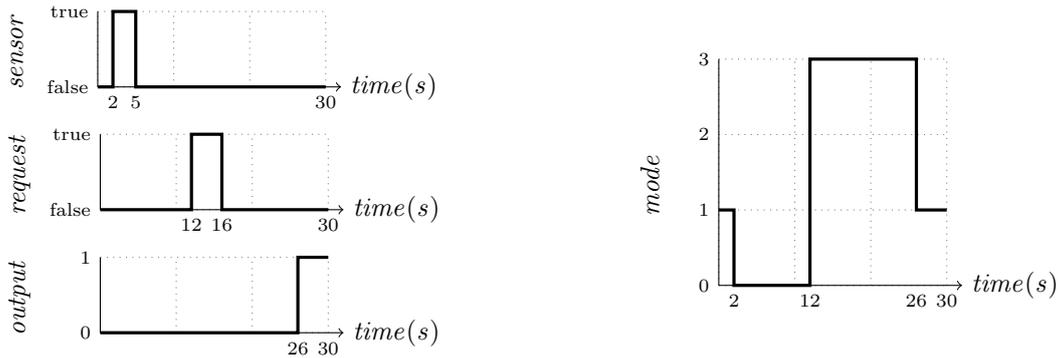

Figure~\ref{fig:vm-signals} illustrates a scenario assuming continuous observation of the input and output signals. If we had chosen to observe the system discretely, we would have a similar scenario, but with a discrete number of samples over time.

In this scenario, a coin is inserted 2s after starting the machine, and a coffee request is performed 10s later. The first input drives the system to the \emph{choice} mode, whereas the second one to the \emph{weak coffee} mode. A weak coffee is produced 14s after the request, which is reflected by changing the machine \emph{output} signal.

An s-DFRS is a 6-tuple: (\emph{I}, \emph{O}, \emph{T}, \emph{gcvar}, \emph{s$_{0}$}, \emph{F}). Inputs (\emph{I}) and outputs (\emph{O}) are system variables, whereas timers (\emph{T}) are used to model temporal behaviour. The global clock is \emph{gcvar}, a variable whose values are non-negative numbers. The initial state is \emph{s$_{0}$}, and \emph{F} is a set of functions describing the system behaviour.

An e-DFRS differs from the symbolic one as it encodes the system behaviour as a state-based machine, whereas an s-DFRS does that symbolically via definitions of functions. An e-DFRS represents a timed system with continuous or discrete behaviour modelled as a state-based machine. States are obtained from an s-DFRS by applying its functions to non-stable states (when a system reaction is expected), but also letting the time evolve. Therefore, an e-DFRS is a 7-tuple: (\emph{I}, \emph{O}, \emph{T}, \emph{gcvar}, \emph{s$_{0}$}, \emph{S}, \emph{TR}), where \emph{TR} is a transition relation associating states in $S$ by means of delay and function transitions. A delay transition represents the observation of the input signals' values after a given delay, whereas the function transition represents how the system reacts to the input signals: the observed values of the output signals. The transitions are encoded as assignments to input and output variables as well as timers.

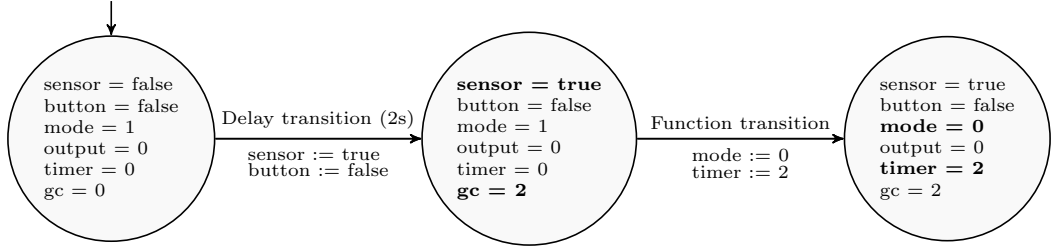
\begin{figure}[h!]
	\centering
	\begin{tikzpicture}[->, >=stealth', auto, semithick, initial where=above, initial text={}, node distance=5.5cm]
	\tikzstyle{every state}=[fill=gray!5,text=black,font=\scriptsize]
	\node[initial,state,align=left] (s1)
	{sensor = false\\ button = false\\ mode = 1\\ output = 0\\ timer = 0\\ gc = 0};
	\node[state,align=left] (s2) [right of=s1]
	{\textbf{sensor = true}\\ button = false\\ mode = 1\\ output = 0\\ timer = 0\\ \textbf{gc = 2}};
	\node[state,align=left] (s3) [right of=s2]
	{sensor = true\\ button = false\\ \textbf{mode = 0}\\ output = 0\\ \textbf{timer = 2}\\ gc = 2};
	
	\path
	(s1)
	edge [above] node[align=left] {\scriptsize{Delay transition (2s)}} (s2)
	edge [below] node[align=left] {\scriptsize{sensor := true}\\[-0.2cm] \scriptsize{button := false}} (s2)
	
	(s2)
	edge [above] node[align=left] {\scriptsize{Function transition}} (s3)
	edge [below] node[align=left] {\scriptsize{mode := 0}\\[-0.2cm] \scriptsize{timer := 2}} (s3);
	\end{tikzpicture}
	\caption{Some states of the e-DFRS representation for the vending machine}
	\label{fig:vm-states}
\end{figure}

Considering the example presented in Figure.~\ref{fig:vm-signals}, Figure.~\ref{fig:vm-states} shows some states of the e-DFRS representation for the vending machine. The initial state considers the initial value of all system variables. The delay transition represents the change of the sensor signal from \emph{false} to \emph{true} after elapsing 2 seconds. Note that the value of \emph{sensor} and the system global clock (\emph{gc}) is updated in the state reached by the delay transition. At this moment, a system reaction is expected, which is characterised by a function transition, updating the system mode, besides resetting the request timer. Here, the reset operation is represented as assigning to the timer the current system global clock. The underlying reason is later explained. The function transition happens instantaneously (time does not evolve).

As said before, we refer to~\cite{carvalho:dfrs_journal} for a comprehensive explanation of DFRS models, besides showing how they can be derived from natural-language requirements.

\subsection{The Coq proof assistant}\label{sec:coq}

Opposed to automatic theorem provers, which aim to develop proofs in a full automatic manner, interactive theorem provers (also known as proof assistants) are tools that mix human interaction and some degree of automation when building proofs. Here, we present Coq~\cite{bertot:coqart}, which is used later in this work.

Coq employs a functional language (Gallina), which is similar to Haskell, to describe algorithms. Computer-verified proofs are developed interactively using tactics, which have some limited support for automation via the tactics language (Ltac). As a logic system, Coq considers a higher-order logic. In what follows, we briefly address these three topics (Gallina, tactics, and automation). Moreover, we also explain the benefits and limitations of functional, logic, and inductive definitions in Coq.

\subsubsection{The Gallina language} Gallina is a typical functional language, with support to define new types, besides polymorphic and higher-order functions. When defining a non-recursive function, one should use the keyword \emph{Definition}. See the following example (\emph{is\_empty}) of a polymorphic function (valid for any given type \emph{T}) that yields a logic value (\emph{True} or \emph{False}) indicating whether the given list is empty.
\ \\
	\begin{coqdoccode}
		\coqdocemptyline		
		\coqdocnoindent
		\coqdockw{Definition} \coqdocvar{is\_empty} \{\coqdocvar{T} : \coqdockw{Type}\} (\coqdocvar{l} : \coqdocvar{list} \coqdocvar{T}) : \coqdockw{Prop} :=\coqdoceol
		\coqdocindent{1.00em}
		\coqdockw{match} \coqdocvar{l} \coqdockw{with}\coqdoceol
		\coqdocindent{1.00em}
		\ensuremath{|} []  \ensuremath{\Rightarrow} \coqdocvar{True}\coqdoceol
		\coqdocindent{1.00em}
		\ensuremath{|} \coqdocvar{\_}   \ensuremath{\Rightarrow} \coqdocvar{False}\coqdoceol
		\coqdocindent{1.00em}
		\coqdockw{end}.\coqdoceol
		\coqdocemptyline	
	\end{coqdoccode}
\ 

Recursive functions shall use \emph{Fixpoint}. The following example (\emph{length}) yields the number of elements of a given list \emph{l}. By pattern matching, if $l$ is empty, the function yields 0. Otherwise, it yields the length of the list tail (\emph{tl}) plus 1.
\ \\
	\begin{coqdoccode}		
		\coqdocemptyline
		\coqdocnoindent
		\coqdockw{Fixpoint} \coqdocvar{length} \{\coqdocvar{T} : \coqdockw{Type}\} (\coqdocvar{l} : \coqdocvar{list} \coqdocvar{T}) : \coqdocvar{nat} :=\coqdoceol
		\coqdocindent{1.00em}
		\coqdockw{match} \coqdocvar{l} \coqdockw{with}\coqdoceol
		\coqdocindent{1.00em}
		\ensuremath{|} []      \ensuremath{\Rightarrow} 0\coqdoceol
		\coqdocindent{1.00em}
		\ensuremath{|} \coqdocvar{h} :: \coqdocvar{tl} \ensuremath{\Rightarrow} 1 + \coqdocvar{length} \coqdocvar{tl}\coqdoceol
		\coqdocindent{1.00em}
		\coqdockw{end}.\coqdoceol
		\coqdocemptyline
	\end{coqdoccode}
\ 

Differently from other functional languages, in Coq, all functions must terminate on all inputs. To ensure that, each recursion must structurally decrease some (the same) argument. If the decreasing analysis performed by the tool cannot identify such an argument, the corresponding recursive function is not defined. In the previous example (\emph{length}), the decreasing argument is the list itself; each recursive call is performed on a smaller list.

\subsubsection{Building proofs with tactics} In Coq, proofs are developed with the aid of tactics\footnote{Index of built-in tactics: \url{https://coq.inria.fr/refman/coq-tacindex.html}}. In the following example, we prove that the length of a list obtained by an append is equal to the sum of the lengths of the appended lists. Let \emph{app} be the appending function, the theorem \emph{length\_app} formalises the previous statement.
\ \\
	\begin{coqdoccode}
		\coqdocemptyline
		\coqdocnoindent
		\coqdockw{Theorem} \coqdocvar{length\_app} :\coqdoceol
		\coqdocindent{1.00em}
		\coqdockw{\ensuremath{\forall}} (\coqdocvar{T} : \coqdockw{Type}) (\coqdocvar{l1} \coqdocvar{l2} : \coqdocvar{list} \coqdocvar{T}),	\coqdocvar{length} (\coqdocvar{app} \coqdocvar{l1} \coqdocvar{l2}) = \coqdocvar{length} \coqdocvar{l1} + \coqdocvar{length} \coqdocvar{l2}.\coqdoceol
		\coqdocnoindent
		\coqdockw{Proof}.\coqdoceol
		\coqdocindent{1.00em}
		\coqdoctac{intros}. \coqdoctac{induction} \coqdocvar{l1}.\coqdoceol
		- \coqdoctac{simpl}. \coqdoctac{reflexivity}.\coqdoceol
		- \coqdoctac{simpl}. \coqdoctac{rewrite} \coqdocvar{IHl1}. \coqdoctac{reflexivity}.\coqdoceol
		\coqdocnoindent
		\coqdockw{Qed}.\coqdoceol
		\coqdocemptyline
	\end{coqdoccode}
\ 

The tactics employed modifies the proof goal in order to demonstrate its truth. Table~\ref{tab:proof} shows how the proof goal evolves after processing each tactic. The command {\tt Proof.} starts the proof environment, loading the proof goal. After that, {\tt intros.} performs universal instantiation, in order to prove the goal for arbitrary values of {\tt T}, {\tt l1}, and {\tt l2}. The command {\tt induction l1.} performs induction on \emph{l1}. This creates two subgoals: base case, and inductive step. The symbol {\tt -} (optional) tells Coq that we now focus on the next subgoal (base case).

\begin{table}[h!]
	\centering
	\caption{Proof of theorem \emph{length\_app}}
	\label{tab:proof}
	\begin{tabular}{|l|p{9.5cm}|} \hline
		\textbf{Command} & \textbf{Proof state} \\ \hline
		{\tt Proof.} & 
		{\tt forall (T : Type) (l1 l2 : list T),\newline
			length (app l1 l2) = length l1 + length l2}\\ \hline
		{\tt intros.} &
		{\tt length (app l1 l2) = length l1 + length l2}\\ \hline
		{\tt induction l1.} & 
		{\tt \noindent\rule{7cm}{0.3pt}(1/2)\newline
			length (app [] l2) = length [] + length l2\newline
			\noindent\rule{7cm}{0.3pt}(2/2)\newline
			length (app (a :: l1) l2) = length (a :: l1) + length l2}\\ \hline
		{\tt -} &
		{\tt length (app [] l2) = length [] + length l2} \\ \hline
		{\tt simpl.} & 
		{\tt length l2 = length l2} \\ \hline
		{\tt reflexivity.} & 
		{\tt This subproof is complete} \\ \hline
		{\tt -} & 
		{\tt IHl1 : length (app l1 l2) = length l1 + length l2\newline
			\noindent\rule{7cm}{0.3pt}\newline
			length (app (a :: l1) l2) = length (a :: l1) + length l2} \\ \hline
		{\tt simpl.} &
		{\tt IHl1 : length (app l1 l2) = length l1 + length l2\newline
			\noindent\rule{7cm}{0.3pt}\newline 
			S (length (app l1 l2)) = S (length l1 + length l2)} \\ \hline
		{\tt rewrite IHl1.} & 
		{\tt IHl1 : length (app l1 l2) = length l1 + length l2\newline
			\noindent\rule{7cm}{0.3pt}\newline 
			S (length l1 + length l2) =	S (length l1 + length l2)} \\ \hline
		{\tt reflexivity.} & 
		{\tt No more subgoals.} \\ \hline
		{\tt Qed.} &
		{\tt length\_app is defined} \\ \hline
	\end{tabular}
\end{table}

After simplifying the proof goal of the base case ({\tt simpl.}), we are left to prove that {\tt length l2 = length l2}, which is trivially true (proved by {\tt reflexivity.}). In the inductive step, assuming the hypothesis {\tt IHl1 : length (app l1 l2) = length l1 + length l2}, we need to prove that {\tt length (app (a :: l1) l2) = length (a :: l1) + length l2} holds. After simplification ({\tt simpl.}), the goal becomes: {\tt S (length (app l1 l2)) = S (length l1 + length l2)}.

By rewriting the goal considering \emph{IHl1} ({\tt rewrite IHl1.}) we get {\tt S (length l1 + length l2) = S (length l1 + length l2)}, provable by {\tt reflexivity.}. The command {\tt Qed.} finishes the prove. Each command is verified by Coq, and it can only be applied if the underlying premises for its application are satisfied. This ensures the construction of a computer-verifiable proof.

\subsubsection{Proof automation} Support for proof automation comes as tacticals (higher-order tactics -- i.e., tactics that take other tactics as arguments), user-defined tactics, and some decision procedures. To illustrate some of these, consider the proof that a list is empty if, and only if, its length is 0.
\ \\
\begin{minipage}[t]{0.47\textwidth}
	\centering
		\begin{coqdoccode}
			\coqdocemptyline		
			\coqdocnoindent
			\coqdockw{Theorem} \coqdocvar{empty\_length\_0} :\coqdoceol
			\coqdocindent{1.00em}
			\coqdockw{\ensuremath{\forall}} (\coqdocvar{T} : \coqdockw{Type}) (\coqdocvar{l} : \coqdocvar{list} \coqdocvar{T}),\coqdoceol
			\coqdocindent{2.00em}
			\coqdocvar{is\_empty} \coqdocvar{l} \ensuremath{\leftrightarrow} \coqdocvar{length} \coqdocvar{l} = 0.\coqdoceol
			\coqdocnoindent
			\coqdockw{Proof}.\coqdoceol
			\coqdocindent{1.00em}
			\coqdoctac{intros}. \coqdoctac{destruct} \coqdocvar{l}.\coqdoceol
			\coqdocindent{1.00em}
			- \coqdoctac{split}.\coqdoceol
			\coqdocindent{2.00em}
			+ \coqdoctac{simpl}. \coqdoctac{intro} \coqdocvar{H}. \coqdoctac{reflexivity}.\coqdoceol
			\coqdocindent{2.00em}
			+ \coqdoctac{simpl}. \coqdoctac{intro} \coqdocvar{H}. \coqdoctac{reflexivity}.\coqdoceol
			\coqdocindent{1.00em}
			- \coqdoctac{split}.\coqdoceol
			\coqdocindent{2.00em}
			+ \coqdoctac{simpl}. \coqdoctac{intro} \coqdocvar{H}. \coqdoctac{inversion} \coqdocvar{H}.\coqdoceol
			\coqdocindent{2.00em}
			+ \coqdoctac{simpl}. \coqdoctac{intro} \coqdocvar{H}. \coqdoctac{inversion} \coqdocvar{H}.\coqdoceol
			\coqdocnoindent
			\coqdockw{Qed}.\coqdoceol
			\coqdocemptyline
			\coqdocemptyline
			\coqdocemptyline										
		\end{coqdoccode}
\end{minipage}
\begin{minipage}[t]{0.53\textwidth}
	\centering
		\begin{coqdoccode}
			\coqdocemptyline
			\coqdocnoindent
			\coqdockw{Ltac} \coqdocvar{trivial\_hypo} :=\coqdoceol
			\coqdocindent{1.00em}
			\coqdoctac{try} (\coqdoctac{simpl} ; \coqdoctac{intro} \coqdocvar{H} ; \coqdoctac{reflexivity}).\coqdoceol
			\coqdocemptyline
			\coqdocnoindent
			\coqdockw{Ltac} \coqdocvar{absurd\_hypo} :=\coqdoceol
			\coqdocindent{1.00em}
			\coqdoctac{try} (\coqdoctac{simpl} ; \coqdoctac{intro} \coqdocvar{H} ; \coqdoctac{inversion} \coqdocvar{H}).\coqdoceol
			\coqdocemptyline
			\coqdocnoindent
			\coqdockw{Theorem} \coqdocvar{empty\_length\_0'} :\coqdoceol
			\coqdocindent{1.00em}
			\coqdockw{\ensuremath{\forall}} (\coqdocvar{T} : \coqdockw{Type}) (\coqdocvar{l} : \coqdocvar{list} \coqdocvar{T}),\coqdoceol
			\coqdocindent{2.00em}
			\coqdocvar{is\_empty} \coqdocvar{l} \ensuremath{\leftrightarrow} \coqdocvar{length} \coqdocvar{l} = 0.\coqdoceol
			\coqdocnoindent
			\coqdockw{Proof}.\coqdoceol
			\coqdocindent{1.00em}
			\coqdoctac{intros}. \coqdoctac{destruct} \coqdocvar{l} ; \coqdoctac{split} ;\coqdoceol
			\coqdocindent{1.00em}
			\coqdoctac{repeat} (\coqdocvar{trivial\_hypo} ; \coqdocvar{absurd\_hypo}).\coqdoceol
			\coqdocnoindent
			\coqdockw{Qed}.\coqdoceol
			\coqdocemptyline
			\coqdocemptyline		
			\coqdocemptyline				
		\end{coqdoccode}
\end{minipage}
\ 

In our first try (\emph{empty\_length\_0}), we perform a case analysis on {\tt l} (i.e., empty and non-empty list). Since the goal involves an equivalence ($\leftrightarrow$), we use the tactic {\tt split.} to generate two subgoals considering both sides of the implication. The first two cases (when the list is empty) can be trivially proved by reflexivity. The two others (when the list is not empty) are proved by contradiction ({\tt inversion H}), since we have a contradictory hypothesis {\tt H} in the proof context: a non-empty list is empty, or the length of a non-empty list is 0. As one can see, there is some degree of repetition, and thus room for automation, in this proof.

In our second try (\emph{empty\_length\_0'}), we define two tactics (\emph{trivial\_hypo}, and \emph{absurd\_hypo}), which tries to prove the current goal by reflexivity or contradiction, respectively. Then, the theorem is proved by trying to apply these two tactics many times ({\tt repeat}). The command {\tt ;} applies the following commands to all subgoals, and not only to the next one.

\subsubsection{Functional, logical, and inductive characterisations} Another important aspect of Coq to this work is the possibility to define aspects functionally, logically, or inductively. To give a concrete example, consider the following definitions of whether a number is even. The first definition (\emph{evenb}) is a function that yields true (boolean value) if \emph{n} is even, false otherwise. In this definition, \emph{S} denotes the successor of a natural number. If \emph{n} is the successor of the successor of some number \emph{n'}, to assess whether \emph{n} is even it suffices to assess whether \emph{n'} is even.
\ \\
\begin{minipage}[t]{0.43\textwidth}
	\centering
		\begin{coqdoccode}
			\coqdocemptyline			
			\coqdocnoindent
			\coqdockw{Fixpoint} \coqdocvar{evenb} (\coqdocvar{n} : \coqdocvar{nat}) : \coqdocvar{bool} :=\coqdoceol
			\coqdocindent{1.00em}
			\coqdockw{match} \coqdocvar{n} \coqdockw{with}\coqdoceol
			\coqdocindent{1.00em}
			\ensuremath{|} 0        \ensuremath{\Rightarrow} \coqdocvar{true}\coqdoceol
			\coqdocindent{1.00em}
			\ensuremath{|} 1        \ensuremath{\Rightarrow} \coqdocvar{false}\coqdoceol
			\coqdocindent{1.00em}
			\ensuremath{|} \coqdocvar{S} (\coqdocvar{S} \coqdocvar{n'}) \ensuremath{\Rightarrow} \coqdocvar{evenb} \coqdocvar{n'}\coqdoceol
			\coqdocindent{1.00em}
			\coqdockw{end}.\coqdoceol
			\coqdoceol
		\end{coqdoccode}
\end{minipage}
\begin{minipage}[t]{0.55\textwidth}
	\centering
		\begin{coqdoccode}
			\coqdocemptyline			
			\coqdockw{Definition} \coqdocvar{is\_even} (\coqdocvar{n} : \coqdocvar{nat}) : \coqdockw{Prop} :=\coqdoceol
			\coqdocindent{1.00em}
			\coqdoctac{\ensuremath{\exists}} \coqdocvar{k}, \coqdocvar{n} = \coqdocvar{k} + \coqdocvar{k}.\coqdoceol
			\coqdocemptyline
			\coqdocemptyline				
			\coqdocnoindent
			\coqdockw{Inductive} \coqdocvar{even} : \coqdocvar{nat} \ensuremath{\rightarrow} \coqdockw{Prop} :=\coqdoceol
			\coqdocindent{1.00em}
			\ensuremath{|} \coqdocvar{ev\_0}  : \coqdocvar{even} 0\coqdoceol
			\coqdocindent{1.00em}
			\ensuremath{|} \coqdocvar{ev\_SS} : \coqdockw{\ensuremath{\forall}} \coqdocvar{n}, \coqdocvar{even} \coqdocvar{n} \ensuremath{\rightarrow} \coqdocvar{even} (\coqdocvar{S} (\coqdocvar{S} \coqdocvar{n})).\coqdoceol
			\coqdoceol	
		\end{coqdoccode}
\end{minipage}
\ 

The second definition (\emph{is\_even}) defines this concept in logical terms: a natural number \emph{n} is even if, and only if, there is a natural number \emph{k}, such that $n = k + k$. The third, and last, definition (\emph{even}) characterises this concept inductively. The definitions \emph{ev\_0} and \emph{ev\_SS} can be seen as inference rules, stating that 0 is even (\emph{ev\_0}), and that if \emph{n} is even, the successor of its successor is also even (\emph{ev\_SS}).

Although these three definitions properly capture the concept of being even, proving facts using these definitions might differ significantly. For the last two definitions (logical and inductive ones), one will need to use specific tactics to deal with the existential quantifier and the inference rules, respectively. Differently, concerning the functional definition, one can use the tactic {\tt simpl.} to simplify the proof goal by evaluating the function \emph{evenb} for the given arguments. However, in some situations, due to the termination requirement of Coq for functions, one cannot rely on a purely functional definition.

In this work, when dealing with concrete examples of DFRSs, we favor their functional characterisation to enable automatic proof of model consistency.

\subsection{Property-based testing}\label{sec:pbt}

Testing is an extremely important task for software development, also complimentary to proofs. Even in the presence of proved components, we typically need to integrate them to unproved ones, and thus we need to rely on testing to analyse integration. Additionally, testing can be used as a quick tool to evaluate properties, before trying to prove them. If we submit a property to a large and relevant number of test cases, and it does not fail, we get confidence on its correctness. If it fails, we save proof effort on trying to prove {\tt False}.

Property-based testing, famous in the functional world due to the QuickCheck framework for Haskell~\cite{claessen:quickcheck}, consists of random generation of input data in order to test a computable (executable) property. It comprises four ingredients: (i) an executable property $P$, (ii) generators of random input values for $P$, (iii), printers for reporting counterexamples, and (iv) shrinkers to minimise counterexamples.

A simple example shown in the QuickCheck manual\footnote{QuickCheck: \url{http://www.cse.chalmers.se/~rjmh/QuickCheck/manual.html}} describes how to test whether the reverse of the reverse of a list is equal to the original list. First, one needs to define this property in Haskell:

\begin{flushleft}
	\textit{prop\_RevRev xs = reverse (reverse xs) == xs \\
		where types = xs::[Int]}
\end{flushleft}

Then, QuickCheck is called to try to falsify the property. In this case, no counterexample is found, which is expected, since the property is actually true. However, if testing whether the reverse of a list is equal to the original list, a counterexample should be easily found, and presented to the user.

The concept of property-based testing is supported in Coq via the QuickChick tool, which is an adaptation of QuickCheck ideas for the Coq proof assistant. In this work, we use the QuickChick tool for generating test cases for DFRSs.

\section{Characterisation of symbolic DFRSs in Coq}\label{sec:sdfrs_coq}

For a comprehensive explanation of DFRSs (using Z), a discussion of their expressiveness, along with theoretical and practical validations, besides an explanation of how such models can be automatically derived from controlled natural-language requirements, we refer to~\cite{carvalho:dfrs_journal}. Here, we highlight some aspects of our Coq representation, which main benefit is to have a single computer-verifiable framework for specification, verification and testing. It also enables the extraction of functional code (Haskell/Ocaml) directly from the verified and formal functional definitions; contributing to the development of correct-by-construction tools.

\begin{figure}[t]
	\centering
	\includegraphics[width=\textwidth]{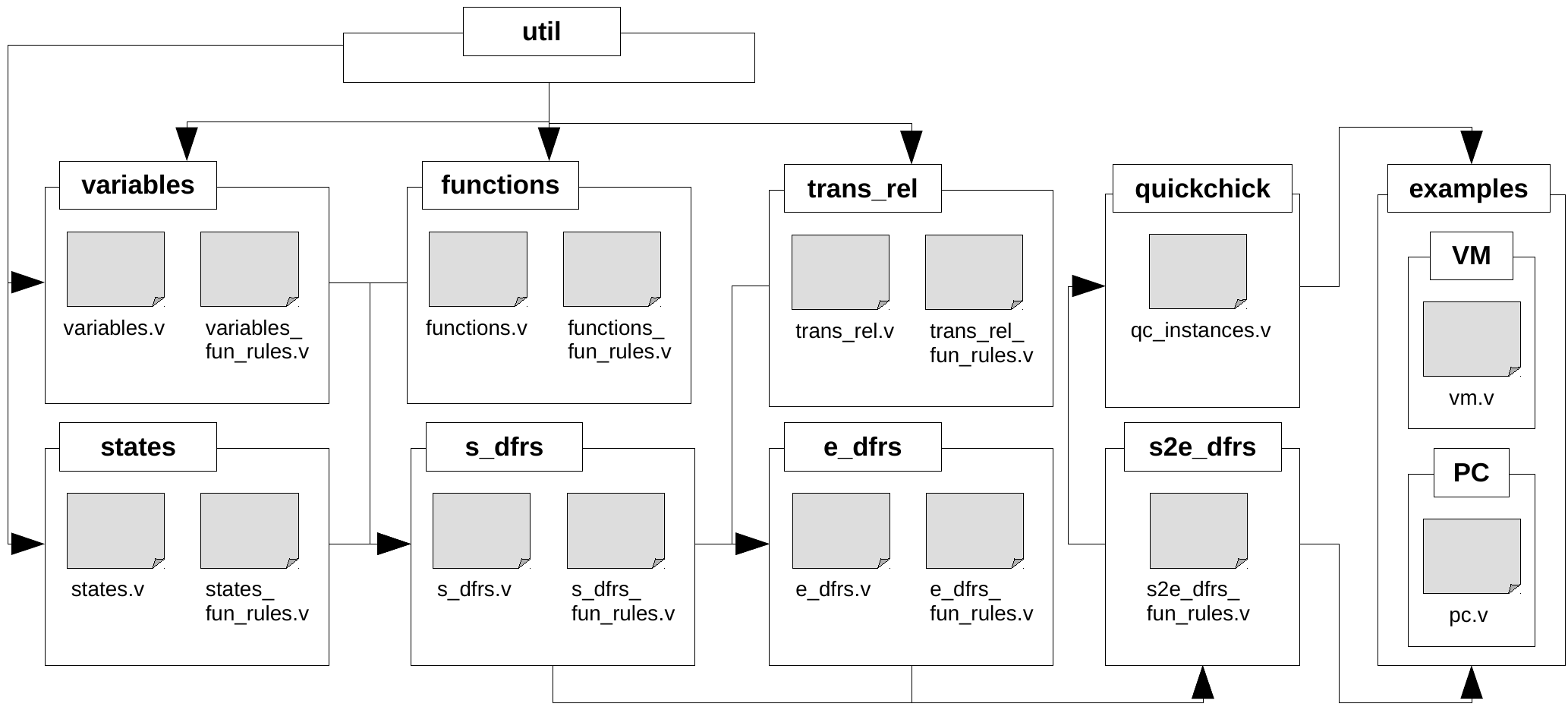}
	\caption{Characterisation of DFRSs in Coq}
	\label{fig:architecture}
\end{figure}

The general architecture of our characterisation of DFRSs in Coq\footnote{\label{fot:git}Available online in: \url{https://github.com/igormeira/DFRScoq}} is presented in Figure~\ref{fig:architecture}. The folders \emph{variables}, \emph{states}, and \emph{functions} define the constituent elements of a symbolic DFRS (\emph{s\_dfrs}). The folder \emph{e\_dfrs} contains the definitions of an expanded DFRS, which is built upon the symbolic one and the definition of a transition relation (\emph{trans\_rel}). Each folder has a main *.v file (named after the folder's name), which provides a logical characterisation. There are two other files: *\_fun\_rules.v (a functional characterisation), and *\_fun\_ind\_equiv.v (proving that both characterisations are equivalent -- not shown in Figure~\ref{fig:architecture} to save space). These proofs allow us to create instances of DFRSs, proving automatically that they are consistent. This aspect is further explained later.

In \emph{s2e\_dfrs}, we have functions that allow for a dynamic expansion of an e\_DFRS from a given s\_DFRS. Part of these functions are used by our test generation module (defined in \emph{quickchick}). Finally, the examples considered in this paper are defined in \emph{examples}. One might rely on the \emph{quickchick} module to generate test cases or on the \emph{s2e\_dfrs} module to perform bounded expansion of an \emph{e\_DFRS}. In what follows, we detail our Coq characterisation of DFRSs.

\subsection{Logical characterisation of symbolic DFRSs}\label{sec:logical_dfrs}

As said before, an s\_DFRS is a 6-tuple: (\emph{I}, \emph{O}, \emph{T}, \emph{gcvar}, \emph{s$_{0}$}, \emph{F}), comprising (input, output, and timer) variables, besides a global clock, along with an initial state, and a set of functions describing the system behaviour.

\subsubsection{Variables}

System variables (\emph{SVARS}), which might represent input or output variables, are defined as a list of pairs of names and types (\emph{TYPE}). A variable name (\emph{VNAME}) is basically a string (\emph{NAME} is defined as a string) that cannot be the reserved name of the system global clock, which is ``gc''.
\ \\
\begin{minipage}[t]{0.45\textwidth}
		\begin{coqdoccode}
			\coqdocemptyline
			\coqdocnoindent
			\coqdockw{Definition} \coqdocvar{ind\_rules\_svars}	(\coqdocvar{svars} :\coqdoceol
			\coqdocnoindent\coqdocvar{list} (\coqdocvar{VNAME} \ensuremath{\times} \coqdocvar{TYPE})) : \coqdockw{Prop} :=\coqdoceol
			\coqdocindent{1.00em}
			\coqdocvar{is\_function} (\coqdocvar{map} (\coqdockw{fun} \coqdocvar{p} \ensuremath{\Rightarrow} \coqdocvar{fst} \coqdocvar{p})\coqdoceol
			\coqdocindent{6.50em}\coqdocvar{svars}) \coqdocvar{comp\_vname}\coqdoceol
			\coqdocindent{1.00em}
			\ensuremath{\land} \ensuremath{\lnot} (\coqdocvar{length} \coqdocvar{svars} = 0)\coqdoceol
			\coqdocindent{1.00em}
			\ensuremath{\land} \coqdocvar{ind\_svars\_valid\_type} \coqdocvar{svars}.\coqdoceol		
			\coqdocemptyline			
		\end{coqdoccode}
\end{minipage}
\begin{minipage}[t]{0.49\textwidth}
		\begin{coqdoccode}
			\coqdocemptyline
			\coqdocnoindent
			\coqdockw{Record} \coqdocvar{VNAME} : \coqdockw{Set} := \coqdocvar{mkVNAME} \{\coqdoceol
			\coqdocindent{1.00em}
			\coqdocvar{vname} : \coqdocvar{NAME}\coqdoceol
			\coqdocindent{1.00em}
			; \coqdocvar{rules\_vname} : \coqdocvar{ind\_rules\_vname} \coqdocvar{vname}\coqdoceol
			\coqdocnoindent
			\}.\coqdoceol
			\coqdocnoindent
			\coqdockw{Record} \coqdocvar{SVARS} : \coqdockw{Set} := \coqdocvar{mkSVARS} \{\coqdoceol
			\coqdocindent{1.00em}
			\coqdocvar{svars} : \coqdocvar{list} (\coqdocvar{VNAME} \ensuremath{\times} \coqdocvar{TYPE})\coqdoceol
			\coqdocindent{1.00em}
			; \coqdocvar{rules\_svars} : \coqdocvar{ind\_rules\_svars} \coqdocvar{svars}\coqdoceol
			\coqdocnoindent
			\}.\coqdoceol		
			\coqdocemptyline
			\coqdocemptyline
			\coqdocemptyline					
		\end{coqdoccode}
\end{minipage}
\

It is important to note the usage of {\tt Record}. In Coq, differently from typical programming languages, a record might comprise data values (\emph{vnames}  and \emph{svars}), but also properties (predicates) that need to hold (\emph{rules\_vname} and \emph{rules\_svars}). Therefore, when creating an instance of a record, it is necessary to prove that the corresponding properties hold. This feature brings cohesion between structural definition and consistency properties.

For the \emph{VNAME} record, the property \emph{rules\_vname} enforces that the variable name (\emph{vname}) is different from ``gc''. Concerning \emph{SVARS}, we have that \emph{svars} needs to be a non-empty function of variable names to types (no name repetition, and each name is mapped to a single type). Besides that, we need to have type consistency. These properties are defined with the aid of propositional functions (i.e., functions that build predicates from the given arguments): \emph{ind\_rules\_vname} (not shown here), and \emph{ind\_rules\_svars}. A record instance is created as follows.
\ \\
	\begin{coqdoccode}
		\coqdocemptyline
		\coqdocnoindent		
		\coqdockw{Definition} \coqdocvar{the\_coin\_sensor} : \coqdocvar{VNAME}.\coqdoceol
		\coqdocnoindent
		\coqdockw{Proof}.\coqdoceol
		\coqdocindent{1.00em}
		\coqdoctac{apply} (\coqdocvar{mkVNAME} ``the\_coin\_sensor'').\coqdoceol
		\coqdocindent{1.00em}\coqdoctac{unfold} \coqdocvar{ind\_rules\_vname}, \coqdocvar{gc}, \coqdocvar{not}. \coqdoctac{intro} \coqdocvar{H}. \coqdoctac{inversion} \coqdocvar{H}.\coqdoceol
		\coqdocnoindent
		\coqdockw{Defined}.\coqdoceol
		\coqdocemptyline
	\end{coqdoccode}
\

First, we give a name for the instance (\emph{the\_coin\_sensor}), and then Coq enters on proof mode. After providing the value for the instance (``the\_coin\_sensor'') via the command {\tt apply (\emph{mkVNAME} ...).}, we need to prove that the defined properties hold for the given value; in this case, that ``the\_coin\_sensor'' is not equal to ``gc''. By unfolding the definitions of \emph{ind\_rules\_vname}, \emph{gc} and \emph{not} we are left to prove that {\tt string\_dec} \emph{``the\_coin\_sensor'' ``gc''} $\rightarrow$ {\tt False}. In Coq, the logical negation ($\lnot P$) is modelled as $P \rightarrow {\tt False}$. Then, we move the antecedent of the implication to the proof context ({\tt intro H.}), and finish the proof since {\tt H} is a contradiction ({\tt inversion H.}). This feature (records with values and predicates) prevents us from creating inconsistent instances (violating rules).

It is worth noting that the proof is finished with {\tt Defined.} instead of {\tt Qed.}. This makes the definition transparent, and it can be unfolded later (we will be able to retrieve the string associated with \emph{vname}). When a proof is finished with {\tt Qed.}, it is marked as opaque (proof irrelevance). To model system timers, we define a different record \emph{STIMERS}. DFRS variables are defined as follows:
\ \\
	\begin{coqdoccode}
		\coqdocemptyline
		\coqdocnoindent
		\coqdockw{Record} \coqdocvar{DFRS\_VARIABLES} : \coqdockw{Set} := \coqdocvar{mkDFRS\_VARIABLES} \{\coqdoceol
		\coqdocindent{1.00em}
		\coqdocvar{I} : \coqdocvar{SVARS} ;	\coqdocvar{O} : \coqdocvar{SVARS} ; \coqdocvar{T} : \coqdocvar{STIMERS} ; \coqdocvar{gcvar} : \coqdocvar{NAME} \ensuremath{\times} \coqdocvar{TYPE}\coqdoceol
		\coqdocindent{1.00em}
		; \coqdocvar{rules\_dfrs\_variables} : \coqdocvar{ind\_rules\_dfrs\_variables} \coqdocvar{I} \coqdocvar{O} \coqdocvar{T} \coqdocvar{gcvar}\coqdoceol
		\coqdocnoindent
		\}.\coqdoceol
	\end{coqdoccode}	
\ 

The definition of \emph{ind\_rules\_dfrs\_variables} guarantees that: (i) the name of the \emph{gc} variable is the string ``gc'', (ii) \emph{I}, \emph{O}, and \emph{T} are disjoint (different names), and (iv) we have type consistency between timers and the global clock (they share the same type). For the vending machine, we have the following definitions.
\ \\
	\begin{coqdoccode}
		\coqdocemptyline
		\coqdocnoindent
		\coqdockw{Definition} \coqdocvar{vm\_I} : \coqdocvar{SVARS}. \coqdockw{Proof}.\coqdoceol
		\coqdocindent{1.00em}\coqdoctac{apply} (\coqdocvar{mkSVARS} [(\coqdocvar{the\_coin\_sensor}, \coqdocvar{Tbool}) ; (\coqdocvar{the\_coffee\_request\_button}, \coqdocvar{Tbool})]).\coqdoceol
		\coqdocindent{1.00em}\emph{(* proof omitted *)} \coqdockw{Defined}.\coqdoceol	
		\coqdocemptyline
		\coqdocnoindent
		\coqdockw{Definition} \coqdocvar{vm\_variables} : \coqdocvar{DFRS\_VARIABLES}. \coqdockw{Proof}.\coqdoceol
		\coqdocindent{1.00em}\coqdoctac{apply} (\coqdocvar{mkDFRS\_VARIABLES} \coqdocvar{vm\_I} \coqdocvar{vm\_O} \coqdocvar{vm\_T} \coqdocvar{vm\_gcvar}). \emph{(* proof omitted *)}\coqdoceol
		\coqdocnoindent
		\coqdockw{Defined}.\coqdoceol
		\coqdocemptyline
	\end{coqdoccode}	
\ 

The element \emph{vm\_I} defines the input variables; \emph{vm\_O}, \emph{vm\_T}, and \emph{vm\_gcvar} are analogous. Note that the element \emph{the\_coin\_sensor} was defined in a previous example.

\subsubsection{Initial state}

A state is a list of names mapped to a pair of values. See the following definitions.
\ \\
	\begin{coqdoccode}
		\coqdocemptyline
		\coqdocnoindent
		\coqdockw{Record} \coqdocvar{STATE} : \coqdockw{Set} := \coqdocvar{mkSTATE} \{\coqdoceol
		\coqdocindent{1.00em}
		\coqdocvar{state} : \coqdocvar{list} (\coqdocvar{NAME} \ensuremath{\times} (\coqdocvar{VALUE} \ensuremath{\times} \coqdocvar{VALUE}))\coqdoceol
		\coqdocindent{1.00em}
		; \coqdocvar{rules\_state} : \coqdocvar{ind\_rules\_state} \coqdocvar{state} \}.\coqdoceol
		\coqdocemptyline
		\coqdocnoindent
		\coqdockw{Record} \coqdocvar{DFRS\_INITIAL\_STATE} : \coqdockw{Set} :=\coqdoceol
		\coqdocindent{1.00em} \coqdocvar{mkDFRS\_INITIAL\_STATE} \{ \coqdocvar{s0} : \coqdocvar{STATE} \}.\coqdoceol
		\coqdocemptyline
	\end{coqdoccode}
\ 

The pair elements are the previous/current variable values. The property \emph{ind\_rules\_state} enforces that \emph{state} is also a function. The initial state for the VM is: both input signals are false, the system mode is idle (\emph{i 1}), the machine output is strong coffee (\emph{i 0}), and the timer and the global clock are equal to 0.
\ \\
	\begin{coqdoccode}
		\coqdocemptyline
		\coqdocnoindent
		\coqdockw{Definition} \coqdocvar{vm\_state} : \coqdocvar{STATE}.\coqdoceol
		\coqdocnoindent
		\coqdockw{Proof}. \coqdoctac{apply} (\coqdocvar{mkSTATE}\coqdoceol
		\coqdocindent{7.00em}
		[(``the\_coin\_sensor'', (\coqdocvar{b} \coqdocvar{false}, \coqdocvar{b} \coqdocvar{false}));\coqdoceol
		\coqdocindent{7.20em}
		(``the\_coffee\_request\_button'', (\coqdocvar{b} \coqdocvar{false}, \coqdocvar{b} \coqdocvar{false}));\coqdoceol
		\coqdocindent{7.20em}
		(``the\_system\_mode'', (\coqdocvar{i} 1, \coqdocvar{i} 1));\coqdoceol
		\coqdocindent{7.20em}
		(``the\_coffee\_machine\_output'', (\coqdocvar{i} 0, \coqdocvar{i} 0));\coqdoceol
		\coqdocindent{7.20em}
		(``the\_request\_timer'', (\coqdocvar{n} 0, \coqdocvar{n} 0)); (``gc'', (\coqdocvar{n} 0, \coqdocvar{n} 0))]).\coqdoceol
		\coqdocindent{1.00em}
		\emph{(* proof omitted *)}\coqdoceol
		\coqdocnoindent
		\coqdockw{Defined}.\coqdoceol
	\end{coqdoccode} 

\subsubsection{Functions}

A DFRS might comprise multiple concurrent components. The behaviour of each component is described by a function. The behaviour of the entire s\_DFRS is then defined as a list of functions (\emph{F}), which cannot be empty (ensured by \emph{ind\_rules\_dfrs\_functions}). Each function (\emph{function}) is a list of 4-tuples: a static guard, a timed guard, a list of assignments, and requirement traceability information. The first two elements define the static and timed conditions necessary to be met to react by performing the respective assignments. One of these two conditions can be empty, but not both (ensured by \emph{ind\_rules\_function}).
\ \\
	\begin{coqdoccode}
		\coqdocemptyline
		\coqdocnoindent
		\coqdockw{Record} \coqdocvar{FUNCTION} : \coqdockw{Set} := \coqdocvar{mkFUNCTION} \{\coqdoceol
		\coqdocindent{1.00em}
		\coqdocvar{function} : \coqdocvar{list} (\coqdocvar{EXP} \ensuremath{\times} \coqdocvar{EXP} \ensuremath{\times} \coqdocvar{ASGMTS} \ensuremath{\times} \coqdocvar{REQUIREMENT}) ;\coqdoceol
		\coqdocindent{1.00em}
		\coqdocvar{rules\_function} : \coqdocvar{ind\_rules\_function} \coqdocvar{function}	\}.\coqdoceol
		\coqdocemptyline
		\coqdocnoindent
		\coqdockw{Record} \coqdocvar{DFRS\_FUNCTIONS} : \coqdockw{Set} := \coqdocvar{mkDFRS\_FUNCTIONS} \{\coqdoceol
		\coqdocindent{1.00em}
		\coqdocvar{F} : \coqdocvar{list} \coqdocvar{FUNCTION} ; \coqdocvar{rules\_dfrs\_functions} : \coqdocvar{ind\_rules\_dfrs\_functions} \coqdocvar{F} \}.\coqdoceol
		\coqdocemptyline
	\end{coqdoccode}
\ 

The aforementioned requirement of the VM (REQ001) says that ``when the system mode is idle, and the coin sensor changes to true, the coffee machine system shall: reset the request timer, assign choice to the system mode''. Part of the formalisation of this requirement is shown below: \emph{req001\_sg\_disj3} models one clause of its static guard (the current system mode is equal to idle---1), whereas \emph{req001\_asgmt2} models one of its assignments (updating the system mode to choice---0). The term \emph{DIS} refers to a list of disjunctive clauses; in this example, we have a conjunction of two elements, each one with a single disjunction. In our work, the conditions adhere to a Conjunctive Normal Form (CNF).
\ \\
	\begin{coqdoccode}
		\coqdocemptyline
		\coqdocnoindent
		\coqdockw{Definition} \coqdocvar{req001\_sg\_disj3} : \coqdocvar{DISJ}.\coqdoceol
		\coqdocnoindent
		\coqdockw{Proof}. \coqdoctac{apply} (\coqdocvar{mkDISJ} [\coqdocvar{mkBEXP} (\coqdocvar{current} (\coqdocvar{the\_system\_mode})) \coqdocvar{eq} (\coqdocvar{i} 1)]).\coqdoceol
		\coqdocindent{1.00em}
		\emph{(* proof omitted *)} \coqdockw{Defined}.\coqdoceol
		\coqdocemptyline
		\coqdocnoindent
		\coqdockw{Definition} \coqdocvar{req001\_asgmt2} : \coqdocvar{ASGMT}.\coqdoceol
		\coqdocnoindent
		\coqdockw{Proof}. \coqdoctac{apply} (\coqdocvar{mkASGMT} (\coqdocvar{the\_system\_mode}, (\coqdocvar{i} 0))). 	\coqdockw{Defined}.\coqdoceol
		\coqdocemptyline
	\end{coqdoccode}

\subsubsection{s\_DFRSs}

An s\_DFRS is composed by the previously defined elements. Various consistency properties are enforce by \emph{ind\_rules\_s\_dfrs}; for instance, the initial state must provide values for all system variables. We refer to our git repository for the definition in details of \emph{ind\_rules\_s\_dfrs}.
\ \\
	\begin{coqdoccode}
		\coqdocemptyline
		\coqdocnoindent
		\coqdockw{Record} \coqdocvar{s\_DFRS} : \coqdockw{Set} := \coqdocvar{mkS\_DFRS} \{\coqdoceol
		\coqdocindent{1.00em}
		\coqdocvar{s\_dfrs\_variables}     : \coqdocvar{DFRS\_VARIABLES} ;\coqdoceol
		\coqdocindent{1.00em}
		\coqdocvar{s\_dfrs\_initial\_state} : \coqdocvar{DFRS\_INITIAL\_STATE} ;\coqdoceol
		\coqdocindent{1.00em}
		\coqdocvar{s\_dfrs\_functions}     : \coqdocvar{DFRS\_FUNCTIONS}\coqdoceol
		\coqdocindent{1.00em}
		; \coqdocvar{rules\_s\_dfrs} : \coqdocvar{ind\_rules\_s\_dfrs} \coqdocvar{s\_dfrs\_variables} \coqdocvar{s\_dfrs\_initial\_state} \coqdocvar{s\_dfrs\_functions}\coqdoceol
		\coqdocnoindent
		\}.\coqdoceol
	\end{coqdoccode}

\subsection{Functional characterisation of symbolic/expanded DFRSs}

When defining the element \emph{the\_coin\_sensor}, it was necessary to prove that the variable name is not ``gc''. When more complex consistency rules need to be proved, the proof script becomes equally more complex, which inhibits automation. A workaround consists in providing functionally-defined consistency rules, which are logically equivalent to their logical/inductive counterparts.

The equivalence proof for \emph{ind\_rules\_vname} (a variable name shall be different from ``gc'') is shown below. We prove both sides of the implication ({\tt split.}) by reaching (via different ways) a contradiction in the proof ({\tt inversion H'}).
\ \\
	\begin{coqdoccode}
		\coqdocemptyline
		\coqdocnoindent
		\coqdockw{Theorem} \coqdocvar{theo\_rules\_vname} :\coqdoceol
		\coqdocindent{1.00em}
		\coqdockw{\ensuremath{\forall}} (\coqdocvar{vname} : \coqdocvar{NAME}),	\coqdocvar{ind\_rules\_vname} \coqdocvar{vname} \ensuremath{\leftrightarrow} \coqdocvar{fun\_rules\_vname} \coqdocvar{vname} = \coqdocvar{true}.\coqdoceol
		\coqdocnoindent
		\coqdockw{Proof}.\coqdoceol
		\coqdocindent{1.00em}
		\coqdoctac{intros}. \coqdoctac{unfold} \coqdocvar{fun\_rules\_vname}, \coqdocvar{ind\_rules\_vname}. \coqdoctac{split}.\coqdoceol
		\coqdocindent{1.00em}
		- \coqdoctac{intro} \coqdocvar{H}. \coqdoctac{unfold} \coqdocvar{not} \coqdoctac{in} \coqdocvar{H}.	\coqdoctac{apply} \coqdocvar{eq\_true\_not\_negb}. \coqdoctac{unfold} \coqdocvar{not}. \coqdoctac{intro} \coqdocvar{H'}.\coqdoceol
		\coqdocindent{2.00em}
		\coqdoctac{rewrite} \coqdocvar{theo\_string\_dec} \coqdoctac{in} \coqdocvar{H}.	\coqdoctac{apply} \coqdocvar{H} \coqdoctac{in} \coqdocvar{H'}. \coqdoctac{inversion} \coqdocvar{H'}.\coqdoceol
		\coqdocindent{1.00em}
		- \coqdoctac{intro} \coqdocvar{H}. \coqdoctac{unfold} \coqdocvar{not}. \coqdoctac{intro} \coqdocvar{H'}. \coqdoctac{rewrite} \coqdocvar{negb\_true\_iff} \coqdoctac{in} \coqdocvar{H}.\coqdoceol
		\coqdocindent{2.00em}
		\coqdoctac{rewrite} \coqdocvar{theo\_string\_dec} \coqdoctac{in} \coqdocvar{H'}. \coqdoctac{rewrite} \coqdocvar{H} \coqdoctac{in} \coqdocvar{H'}. \coqdoctac{inversion} \coqdocvar{H'}.\coqdoceol
		\coqdocnoindent
		\coqdockw{Qed}.\coqdoceol
		\coqdocemptyline
	\end{coqdoccode}
\ 

Now it is possible to define \emph{the\_coin\_sensor} as follows. We apply the theorem \emph{theo\_rules\_vname} to rewrite the proof goal considering the functional characterisation. Then, it suffices to execute the tactic {\tt reflexivity.}, which simplifies the goal by performing the necessary computations, besides concluding the proof.
\ \\
	\begin{coqdoccode}
		\coqdocemptyline
		\coqdocnoindent		
		\coqdockw{Definition} \coqdocvar{the\_coin\_sensor} : \coqdocvar{VNAME}.\coqdoceol
		\coqdocnoindent
		\coqdockw{Proof}.\coqdoceol
		\coqdocindent{1.00em}
		\coqdoctac{apply} (\coqdocvar{mkVNAME} ``the\_coin\_sensor''). \coqdoctac{apply} \coqdocvar{theo\_rules\_vname}. \coqdoctac{reflexivity}.\coqdoceol
		\coqdocnoindent
		\coqdockw{Defined}.\coqdoceol
		\coqdocemptyline
	\end{coqdoccode}
\ 

Actually, all proofs omitted in the previous examples follow this pattern. Nevertheless, the logical/inductive characterisation is still useful, mainly when dealing with infinite aspects, which appear on expanded DFRSs. The functional counterpart needs to be bounded to some exploration limit.

\section{Characterisation of expanded DFRSs in Coq}\label{sec:edfrs_coq}

An e\_DFRS is defined in terms of a transition relation (a list of transitions -- \emph{TRANS}). Each transition relates two states by means of a label (\emph{TRANS\_LABEL}). A label denotes a function (\emph{func}) or a delay (\emph{del}) transition. A function transition models system reaction; it changes the value of output variables and timers. A delay transition models time evolving (\emph{DELAY}), besides modifying the value of system inputs. A number of consistency rules are enforced by \emph{rules\_TR}.

\begin{minipage}[t]{0.50\textwidth}
	\centering
		\begin{coqdoccode}
			\coqdocemptyline			
			\coqdocnoindent
			\coqdockw{Inductive} \coqdocvar{TRANS\_LABEL} : \coqdockw{Type} :=\coqdoceol
			\coqdocindent{0.50em}
			\ensuremath{|} \coqdocvar{func}  : (\coqdocvar{ASGMTS} \ensuremath{\times} \coqdocvar{REQUIREMENT})\coqdoceol \coqdocindent{4.00em}\ensuremath{\rightarrow} \coqdocvar{TRANS\_LABEL}\coqdoceol
			\coqdocindent{0.50em}
			\ensuremath{|} \coqdocvar{del}   : (\coqdocvar{DELAY} \ensuremath{\times} \coqdocvar{ASGMTS})\coqdoceol
			\coqdocindent{4.00em}\ensuremath{\rightarrow} \coqdocvar{TRANS\_LABEL}.\coqdoceol
			\coqdocemptyline
			\coqdocnoindent
			\coqdockw{Record} \coqdocvar{TRANS} : \coqdockw{Set} := \coqdocvar{mkTRANS} \{\coqdoceol
			\coqdocindent{0.50em}
			\coqdocvar{STS} : \coqdocvar{STATE} \ensuremath{\times} \coqdocvar{TRANS\_LABEL}\coqdoceol
			\coqdocindent{3.30em}\ensuremath{\times} \coqdocvar{STATE} \}.\coqdoceol
		\end{coqdoccode}
\end{minipage}
\begin{minipage}[t]{0.53\textwidth}
	\centering
		\begin{coqdoccode}
			\coqdocemptyline			
			\coqdocnoindent
			\coqdockw{Record} \coqdocvar{TRANSREL} : \coqdockw{Set} :=\coqdoceol
			\coqdocindent{0.50em}\coqdocvar{mkTRANSREL} \{\coqdoceol
			\coqdocindent{1.00em}
			\coqdocvar{transrel} : \coqdocvar{list} \coqdocvar{TRANS} \}.\coqdoceol		
			\coqdocemptyline
			\coqdocnoindent
			\coqdockw{Record}\coqdoceol
			\coqdocindent{0.20em}\coqdocvar{DFRS\_TRANSITION\_RELATION}\coqdoceol
			\coqdocindent{0.20em}:= \coqdocvar{mkDFRSTRANSITIONREL} \{\coqdoceol
			\coqdocindent{1.00em}
			\coqdocvar{TR} : \coqdocvar{TRANSREL} ;\coqdoceol
			\coqdocindent{1.00em}\coqdocvar{rules\_TR} : \coqdocvar{ind\_rules\_TR TR} ;
			\}.\coqdoceol
		\end{coqdoccode}
\end{minipage}
\ \\

An e\_DFRS is defined as a combination of variables, states, and a transition relation. As said before, an e\_DFRS is obtained by the expansion of the corresponding s\_DFRS, by letting the time evolve (performing delay transitions), and observing how the system reacts to input stimuli (performing function transitions).

\ \\
\begin{coqdoccode}
	\coqdocnoindent
	\coqdockw{Record} \coqdocvar{e\_DFRS} : \coqdockw{Set} := \coqdocvar{mkE\_DFRS} \{\coqdoceol
	\coqdocindent{1.00em}
	\coqdocvar{e\_dfrs\_variables}              : \coqdocvar{DFRS\_VARIABLES};\coqdoceol
	\coqdocindent{1.00em}
	\coqdocvar{e\_dfrs\_states}                 : \coqdocvar{DFRS\_STATES};\coqdoceol
	\coqdocindent{1.00em}
	\coqdocvar{e\_dfrs\_transition\_relation}    : \coqdocvar{DFRS\_TRANSITION\_RELATION};\coqdoceol
	\coqdocnoindent
	\}.\coqdoceol
	\coqdocemptyline
\end{coqdoccode}
\ 

Since time is always expected to be able to evolve, and the system global clock is part of the state, an e\_DFRS comprises an infinite number of states. Therefore, a function that expands a symbolic DFRS, yielding the obtained e\_DFRS, would never terminate its execution and, thus, cannot be defined in Coq. In the following section, we explain how we can perform a bounded construction of an e\_DFRS.

\subsection{Bounded construction of expanded DFRSs}\label{sec:bounded_edfrs}

Typically, an e\_DFRS consists of an infinite number of states, since time might always evolve, reaching a new state (different value for the global clock). Therefore, although one can characterise all states of an e\_DFRS inductively, a functional definition needs to be bounded. Here, we restrict the number of recursive calls of \emph{buildTR}, which expands dynamically an s\_DFRS, up to \emph{num}. When it reaches 0, the function stops. When there are still states to visit (expand), and the limit has not been reached, new states are generated with the aid of the auxiliary function \emph{genTransitions}. Then, the function recurses decreasing the limit by one.
\ \\
	\begin{coqdoccode}	
		\coqdocemptyline
		\coqdocnoindent		
		\coqdockw{Fixpoint} \coqdocvar{buildTR} (\coqdocvar{toVisit} \coqdocvar{visited} : \coqdocvar{list} \coqdocvar{STATE}) (\coqdocvar{I} \coqdocvar{Out} \coqdocvar{T} : \coqdocvar{list} (\coqdocvar{VNAME} \ensuremath{\times} \coqdocvar{TYPE}))\coqdoceol
		\coqdocindent{1.00em}
		(\coqdocvar{F} : \coqdocvar{list} (\coqdocvar{list} \coqdocvar{FUNCTION})) (\coqdocvar{possibilities} : \coqdocvar{list} (\coqdocvar{VNAME} \ensuremath{\times} \coqdocvar{list} \coqdocvar{VALUE}))\coqdoceol
		\coqdocindent{1.00em}
		(\coqdocvar{num} : \coqdocvar{nat}) : \coqdocvar{list} \coqdocvar{TRANS} :=\coqdoceol
		\coqdocindent{1.00em}
		\coqdockw{match} \coqdocvar{toVisit}, \coqdocvar{num} \coqdockw{with}\coqdoceol
		\coqdocindent{1.00em}
		\ensuremath{|} [] , \coqdocvar{\_}    \ensuremath{\Rightarrow} []\coqdoceol
		\coqdocindent{1.00em}
		\ensuremath{|} \coqdocvar{\_} :: \coqdocvar{\_}, 0    \ensuremath{\Rightarrow} []\coqdoceol
		\coqdocindent{1.00em}
		\ensuremath{|} \coqdocvar{h} :: \coqdocvar{t}, \coqdocvar{S} \coqdocvar{n'} \ensuremath{\Rightarrow} \coqdockw{let}	\coqdocvar{tr1} := \coqdocvar{genTransitions} \coqdocvar{h} \coqdocvar{I} \coqdocvar{Out} \coqdocvar{T} \coqdocvar{F} \coqdocvar{possibilities}\coqdoceol
		\coqdocindent{8.00em}
		\coqdoctac{in} \coqdockw{if} \coqdocvar{in\_state\_list} \coqdocvar{h}.(\coqdocvar{variables}) \coqdocvar{visited} \coqdocvar{beq\_state}\coqdoceol
		\coqdocindent{9.40em}
		\coqdockw{then} \coqdocvar{buildTR} \coqdocvar{t} \coqdocvar{visited} \coqdocvar{I} \coqdocvar{Out} \coqdocvar{T} \coqdocvar{F} \coqdocvar{possibilities} \coqdocvar{n'}\coqdoceol
		\coqdocindent{9.40em}
		\coqdockw{else} \coqdocvar{tr1}.(\coqdocvar{transrel}) ++ \coqdoceol
		\coqdocindent{11.80em}
		\coqdocvar{buildTR}	(\coqdocvar{t} ++ (\coqdocvar{get\_list\_states} \coqdocvar{tr1}.(\coqdocvar{transrel}) (\coqdocvar{h} :: \coqdocvar{visited})))\coqdoceol
		\coqdocindent{15.80em}
		(\coqdocvar{h} :: \coqdocvar{visited}) \coqdocvar{I} \coqdocvar{Out} \coqdocvar{T} \coqdocvar{F} \coqdocvar{possibilities} \coqdocvar{n'}\coqdoceol
		\coqdocindent{1.00em}
		\coqdockw{end}.\coqdoceol
		\coqdocemptyline	
	\end{coqdoccode}
\ 

The function \emph{genTransitions} first assesses whether a given state is stable (no system reaction is expected, which means that no static and timed guards evaluate to true). If it is stable, the function generates delay transitions (with the configured time step) considering all possible combinations of input values. Otherwise, the function generates function transitions considering the assignments associated with the static and timed guards that evaluate to true.
\ \\
	\begin{coqdoccode}
		\coqdocemptyline
		\coqdocnoindent
		\coqdockw{Definition} \coqdocvar{genTransitions} (\coqdocvar{s} : \coqdocvar{STATE}) (\coqdocvar{I} \coqdocvar{O} \coqdocvar{T} : \coqdocvar{list} (\coqdocvar{VNAME} \ensuremath{\times} \coqdocvar{TYPE})) \coqdoceol
		\coqdocindent{1.00em}
		(\coqdocvar{F} : \coqdocvar{list} (\coqdocvar{list} \coqdocvar{FUNCTION})) (\coqdocvar{possibilities} : \coqdocvar{list} (\coqdocvar{VNAME} \ensuremath{\times} \coqdocvar{list} \coqdocvar{VALUE})) \coqdoceol
		\coqdocindent{1.00em}
		: \coqdocvar{TRANSREL} :=\coqdoceol
		\coqdocindent{1.00em}
		\coqdockw{let} \coqdocvar{entries} := \coqdocvar{union\_lists} (\coqdocvar{map} (\coqdockw{fun} \coqdocvar{f} : \coqdocvar{FUNCTION} \ensuremath{\Rightarrow} \coqdocvar{f}.(\coqdocvar{function})) (\coqdocvar{union\_lists} \coqdocvar{F})) \coqdoctac{in} \coqdoceol
		\coqdocindent{1.00em}
		\coqdockw{let} \coqdocvar{combinations} := \coqdocvar{gen\_asgmts\_combination} (\coqdocvar{possible\_asgmts} \coqdocvar{possibilities}) [[]] \coqdoctac{in}\coqdoceol
		\coqdocindent{1.00em}
		\coqdockw{if} \coqdocvar{is\_stable} \coqdocvar{s} (\coqdocvar{List.app} \coqdocvar{I} \coqdocvar{O}) \coqdocvar{T} \coqdocvar{F}\coqdoceol
		\coqdocindent{2.40em}
		\coqdockw{then} \coqdocvar{mkTRANSREL} (\coqdocvar{make\_trans\_del} \coqdocvar{s} \coqdocvar{I} \coqdocvar{T} \coqdocvar{combinations})\coqdoceol
		\coqdocindent{2.40em}
		\coqdockw{else} \coqdocvar{mkTRANSREL} (\coqdocvar{make\_trans\_func} \coqdocvar{s} (\coqdocvar{I} ++ \coqdocvar{O}) \coqdocvar{T} \coqdocvar{entries}).\coqdoceol
		\coqdocemptyline
	\end{coqdoccode}	
\ 

This bounded and functional exploration of e\_DFRSs can support the development of simulators for e\_DFRSs. Recall that it is possible to extract Haskell and Ocaml code directly from functional definitions in Gallina. Simulation is an important validation technique, since it enables the analysis of whether the created model properly captures the system's expected behaviour.

\section{Generating test cases with QuickChick}\label{sec:quickchick}

Besides bounded exploration, we also allow for the generation of test data (via QuickChick) by sampling valid traces. To achieve this goal, we define a function (\emph{genValidTrace}) that, given an s\_DFRS, yields random valid traces (a list of transitions). To generate valid sequences of transitions, this function relies upon \emph{genTransitions}, previously defined (see Section~\ref{sec:bounded_edfrs}). From the initial state, it generates the possible transitions. Then, it randomly chooses one possible transition, and calls \emph{genValidTrace} recursively, considering as the current state the one reached by the selected transition. The generation of a trace stops when \emph{size} reaches 0 (similarly to \emph{num} in \emph{buildTR} -- see Section~\ref{sec:bounded_edfrs}), but also when \emph{num} has not reached 0 yet. However, this last situation happens with a lower probability. This is achieved via the operator \emph{freq}.
\ \\
	\begin{coqdoccode}
		\coqdocemptyline
		\coqdocnoindent	
		\coqdockw{Fixpoint} \coqdocvar{genValidTrace} (\coqdocvar{st} : \coqdocvar{STATE}) (\coqdocvar{dfrs} : \coqdocvar{s\_DFRS})\coqdoceol
		\coqdocnoindent
		(\coqdocvar{possibilities} : \coqdocvar{list} (\coqdocvar{VNAME} \ensuremath{\times} \coqdocvar{list} \coqdocvar{VALUE})) (\coqdocvar{size} : \coqdocvar{nat}) : \coqdocvar{G} \coqdocvar{trace} :=\coqdoceol
		\coqdocindent{1.00em}
		\coqdockw{match} \coqdocvar{size} \coqdockw{with}\coqdoceol
		\coqdocindent{1.00em}
		\ensuremath{|} 0       \ensuremath{\Rightarrow} \coqdocvar{ret} []\coqdoceol
		\coqdocindent{1.00em}
		\ensuremath{|} \coqdocvar{S} \coqdocvar{size'} \ensuremath{\Rightarrow} \coqdockw{let}\coqdoceol
		\coqdocindent{2.50em} \coqdocvar{tr} := (\coqdocvar{genTransitions} \coqdocvar{st} \coqdocvar{dfrs}.(\coqdocvar{s\_dfrs\_variables}).(\coqdocvar{I}).(\coqdocvar{svars})\coqdoceol
		\coqdocindent{6.00em}
		\coqdocvar{dfrs}.(\coqdocvar{s\_dfrs\_variables}).(\coqdocvar{O}).(\coqdocvar{svars}) \coqdocvar{dfrs}.(\coqdocvar{s\_dfrs\_variables}).(\coqdocvar{T}).(\coqdocvar{stimers})\coqdoceol
		\coqdocindent{6.00em}
		[\coqdocvar{dfrs}.(\coqdocvar{s\_dfrs\_functions}).(\coqdocvar{F})]	\coqdocvar{possibilities}).(\coqdocvar{transrel})\coqdoceol
		\coqdocindent{1.50em}
		\coqdoctac{in} \coqdocvar{freq} [ (1, \coqdocvar{ret} []) ;\coqdoceol
		\coqdocindent{5.50em}
		(\coqdocvar{size}, \coqdocvar{x}  \ensuremath{\leftarrow} \coqdocvar{next\_label} \coqdocvar{st} \coqdocvar{tr} ;;\coqdoceol
		\coqdocindent{8.00em}
		\coqdocvar{xs} \ensuremath{\leftarrow} \coqdocvar{genValidTrace} (\coqdocvar{nextState} \coqdocvar{st} \coqdocvar{x} \coqdocvar{tr}) \coqdocvar{dfrs} \coqdocvar{possibilities} \coqdocvar{size'} ;; \coqdocvar{ret} (\coqdocvar{x} :: \coqdocvar{xs})) ]\coqdoceol
		\coqdocemptyline	
	\end{coqdoccode}
\ 

When we sample \emph{genValidTrace} with QuickChick, by default, it performs 11 calls to the function. See the QuickChick documentation\footnote{QuickChick manual: \url{softwarefoundations.cis.upenn.edu/qc-current/}} for more information. Table~\ref{tab:tcvm} shows, in a tabular form, a fragment of an output (test data) generated via QuickChick. The labels \emph{time}, \emph{sensor}, \emph{request}, \emph{mode}, and \emph{output} refer to the system global clock, the coin sensor, the coffee request button, the system mode, and the coffee machine output, respectively. For this example, the configured time step was 1 and, thus, we see the system global clock evolving by 1 time unit per test step.

\begin{table}[htb]
	\centering
	\caption{\small{Example of test data generated via QuickChick (VM)}}
	\label{tab:tcvm}
	\normalsize{
		\begin{tabular}{c|c|c|c|c} \hline\hline
			\textbf{time} & \textbf{sensor} & \textbf{request} & \textbf{mode} & \textbf{output} \\ \hline\hline
			0	& false	& false	& 1	& 0	\\ \hline
			1	& false	& false	& 1	& 0	\\ \hline
			2	& true	& true	& 0	& 0	\\ \hline
			3	& true	& true	& 0	& 0	\\ \hline
			4	& false	& false	& 0	& 0	\\ \hline
		\end{tabular}
	}
\end{table}

In this example, a coin is inserted and the coffee request button is pressed both at the time (2 seconds after the test beginning). As expected, the system mode changes to choice (represented as 0). When the system global clock is 4, 2 seconds later, the coin sensor becomes false again, and the coffee request button is released. Although not shown in this tabular representation, as we keep requirement traceability information, when defining functions (see Section~\ref{sec:logical_dfrs} -- Functions), we can also extract requirement coverage information from the generated test data.

\subsection{Considerations on soundness and tool certification}

In order to develop a sound model-based testing theory, typically, it is necessary to consider the following elements: (i) adopt a formal specification language, (ii) assume that it is possible to represent the implementation behaviour using the same language (testability hypothesis), (iii) define an implementation relation expressing correctness of implementations with respect to specification models, (iv) define a test generation and a test execution procedure, and (v) finally prove that these procedures are sound with respect to the implementation relation (i.e., if the execution of a generated test case fails, it means that the implementation under test is not related to the considered specification model by the adopted implementation relation).

In this paper, we use Gallina as a formal specification language, and we use property-based testing tool (QuickChick) to generate test data (input and expected output data). To develop a sound testing theory, it would be necessary to define the missing elements, namely: an implementation relation, a test execution procedure, besides taking into account the details on how QuickChick performs property-based testing. Then, we would have the necessary ingredients to prove the soundness of this Gallina-based testing theory. Regarding the implementation relation, a possible candidate would be a relation similar to the one defined in~\cite{carvalho:csptio}: csptio, a conformance relation for CSP timed input-output conformance relation, which also addressed data-flow reactive systems.

Finally, in some critical domains, tool certification is mandatory prior to its integration into the development process; unless the tool outputs are manually inspected, and evidence is produced in favor of the correctness of them. In our work, we do not expect the integration of the NAT2TEST tool into a development tool chain without manual inspection. Our goal is to aid the verification team by producing test data, but it remains as the team responsibility the analyses of whether the system has been properly tested.

\section{Empirical analyses}\label{sec:empirical}

We evaluate our work by considering the VM and an example from the aerospace domain (provided by Embraer): a priority command function (PC) that decides whether the pilot or copilot side stick will have priority in controlling the airplane. Our evaluation considers performance and quality aspects. All data presented here consider multiple executions. The main threat to validity is external. We cannot generalise the conclusions, since few and not very complex examples were considered. Nevertheless, the results allow interesting insights and provide feasibility evidence for our Coq characterisation and testing strategy.

We generated (multiple times) three datasets: performing 1, 5, and 10 calls to the QuickChick sampling function. Each call performs 11 calls to \emph{genValidTrace}. Therefore, each dataset contained 11, 55, and 110 test cases, respectively. The size of each test is bounded to $size = 100$ (see Section~\ref{sec:quickchick}) -- up to 100 delay/function transitions. The time to generate each dataset is small: VM (1.29s/0.01s; 3.75s/0.27s; 7.03s/0.63s) and PC (1.37s/0.03s; 3.96/0.04s; 7.39s/0.80), $\mu/\sigma$ for each dataset\footnote{Considering an i7 @ 2.40GHz x 4, with 8 GB of RAM running Ubuntu 16.04 LTS.}. The time is linearly proportional to the number of sampling calls.

A mutant-based strength analysis was used to assess the quality of the generated tests. Mutation operators yield a trustworthy comparison of test cases strength because they create erroneous programs in a controlled and systematic way~\cite{andrews:Mutation}. A good test suite should be able to detect the introduced error (kill the mutant). Sometimes, the alive mutant is equivalent to the correct program. In general, this verification is undecidable and too error-prone to be made manually.

\begin{figure}
	\centering
	\begin{subfigure}{.5\textwidth}
		\centering
		\includegraphics[width=\linewidth]{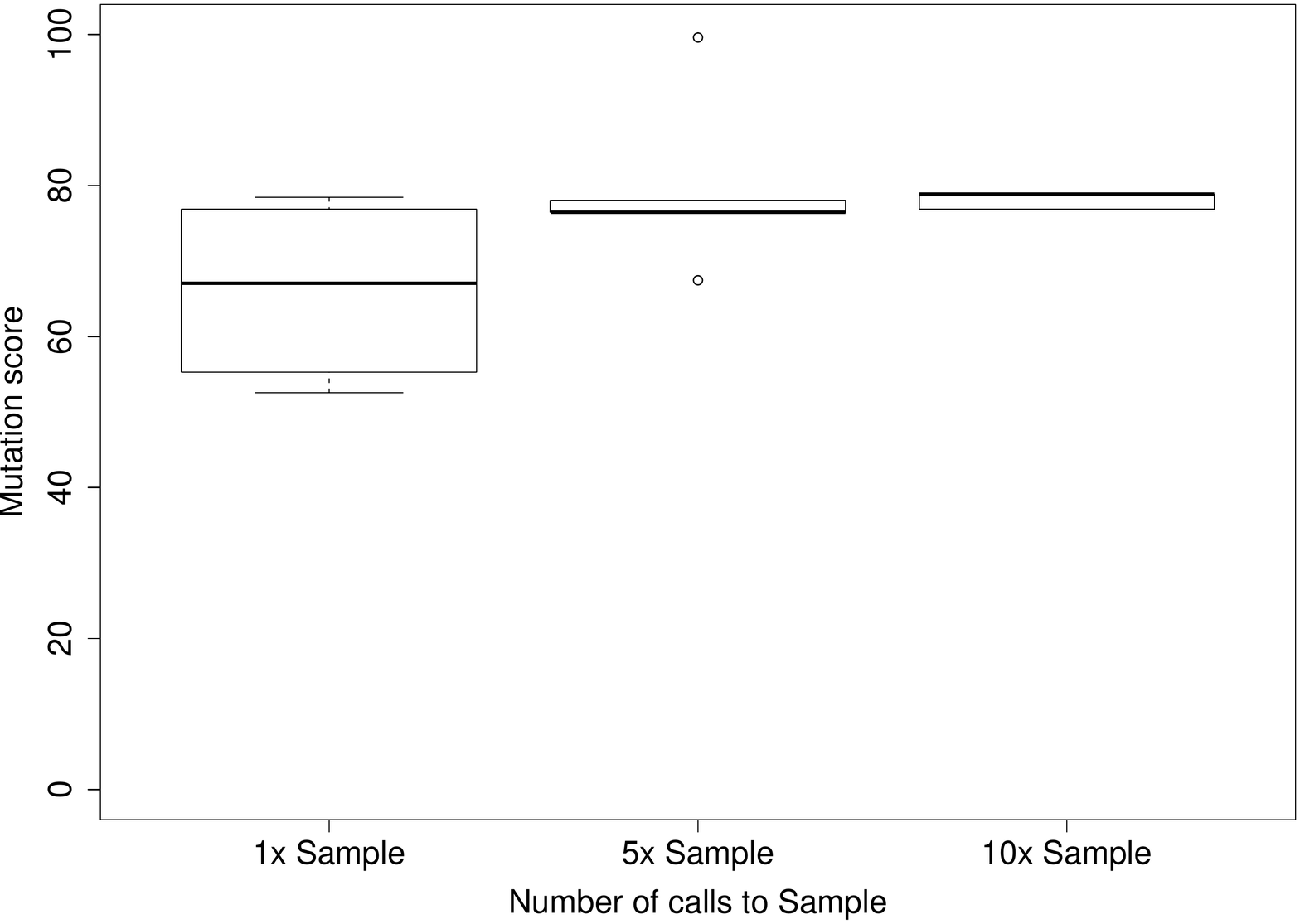}
		\caption{Mutation score for the VM example}
		\label{fig:vm}
	\end{subfigure}%
	\begin{subfigure}{.5\textwidth}
		\centering
		\includegraphics[width=\linewidth]{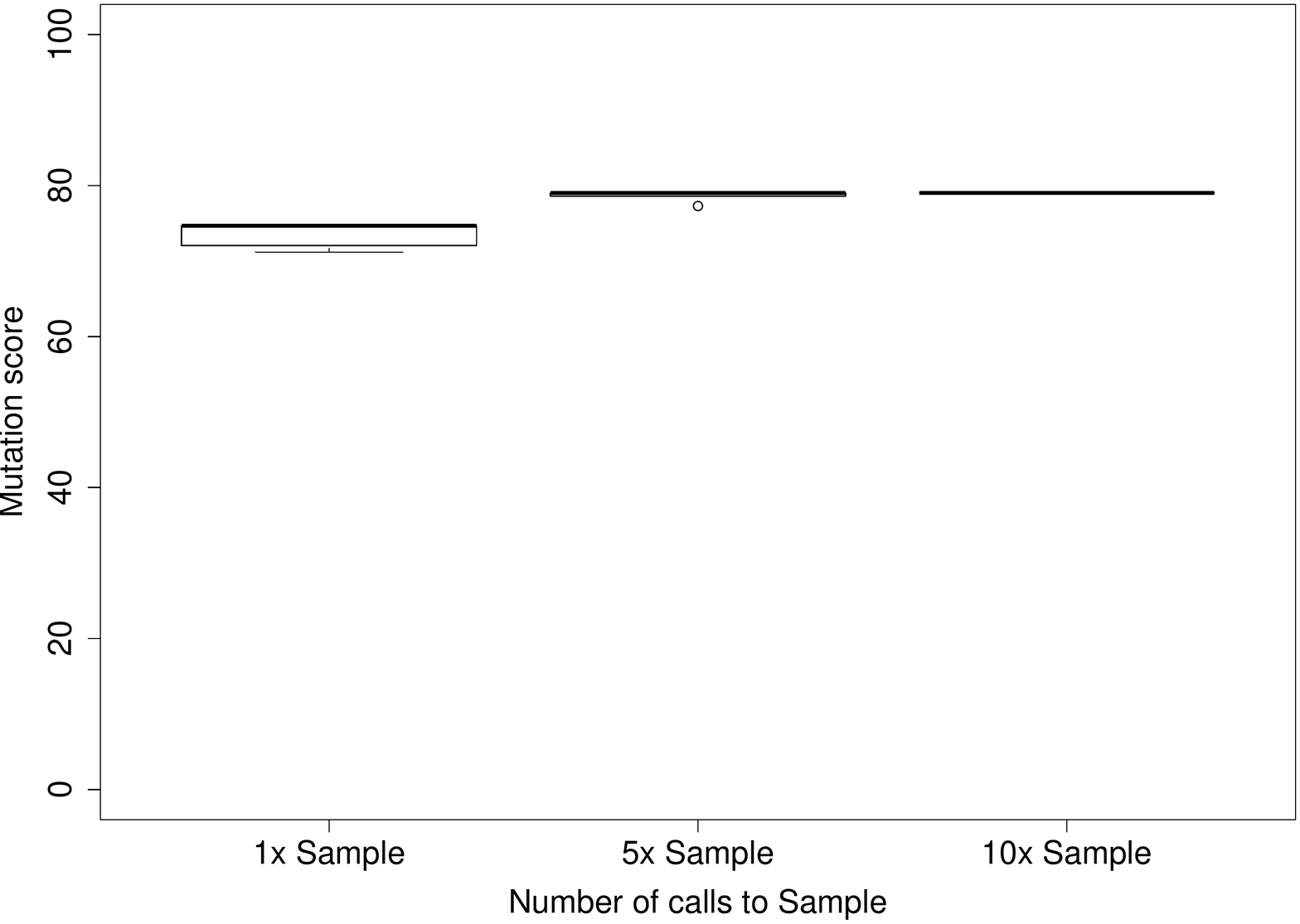}
		\caption{Mutation score for the PC example}
		\label{fig:pc}
	\end{subfigure}
	\caption{Mutant-based strength analysis}
	\label{fig:mbt}
\end{figure}

We follow a conservative approach: all alive mutants are not equivalent ones. This assumption makes the results of the empirical analyses the worst case. We considered C reference implementations, which were mutated via SRCIROR~\cite{hariri:c_mutation}: 255/229 mutants were generated for the VM/PC, respectively. We wrote C test drivers to run all mutants against all generated datasets. Figure~\ref{fig:mbt} shows the mutation score (ratio of killed/generated mutants) for the VM and PC.

In average, the mutation score was 75.80\% (VM: $\mu=74.56\%/\sigma=11.00\%$; PC: $\mu=77.03\%/\sigma=2.81\%$). For the VM, an outlier dataset killed all but one mutant. We inspected the alive one, and it is equivalent to the original code. For the PC, it is worth noting that 5 and 10 calls to the sampling function yielded very similar results. We believe these results are promising, considering the observed performance, besides being fully automatic\footnote{All empirical data and scripts for the VM are available online: see Footnote~\ref{fot:git}. The files regarding the PC example cannot be made publicly available.}. Higher scores should be pursued by complementing the dataset with specialist-defined test scenarios.

\section{Conclusion}\label{sec:conclusion}

This paper presents a logical and a functional characterisation of timed reactive systems in Coq, which can be automatically derived from natural-language requirements. This allows for a single and consistent framework for specifying, verifying, and testing such systems. Moreover, it contributes to the development of correct-by-construction tools, since it is possible to extract Haskell or Ocaml code from the verified and formal functional definitions. Furthermore, tests are automatically generated with the QuickChick tool. Empirical analyses, considering examples both from the literature and the industry, showed that our testing strategy is fast, and can detect about 75.80\% defects introduced by mutation.

The integration of (interactive) theorem provers and testing strategies is a relevant research topic, and thus has been developed by many researches. Similarly to Coq, other provers provide integration between proofs and tests. For instance, considering the theorem prover Isabelle/HOL, we refer to~\cite{brucker:tgen_hol,brucker:theorem_based_testing,nipkow:tgen_isabelle}. In general, similar functionalities are provided among these different options.
Modelling and/or testing timed systems using (interactive) theorem provers is addressed, for example, in~\cite{hong:connectors,brucker:IPC,wan:PLC,ahrendt:tgen_key}, considering timed connectors, real-time operating systems, programmable logic controllers, and Java code, respectively. Differently, our work focus on models of system-level requirements. In this direction, we have the works reported in~\cite{mohring:TA,feliachi:circus}, which consider as modelling notation timed automata and Circus, respectively. However, these models are not derived from natural-language requirements, which is the case here.

\subsection{Future work}

All proofs regarding the equivalence of logical and functional characterisations of DFRS models have been completed. The generation of Coq specifications and test generation via QuickChick have been integrated into the NAT2TEST tool. Furthermore, more empirical analyses have been carried out. These new results will be fully presented and explained in the final version of this paper.

Besides these actions, which have already been carried out, we also plan to support specialist-defined test scenarios. One will be able to define fragments of a test scenario, and, with the aid of QuickChick, we will find a valid trace that includes the user-defined scenario. This will allow for complementing the testing campaign with scenarios that need to be tested, which might not be necessarily covered by a random test-generation strategy.



\end{document}